\documentclass[sigconf]{acmart}

\AtBeginDocument{%
  }

\copyrightyear{2026}
\acmYear{2026}
\setcopyright{cc}
\setcctype{by}
\acmConference[KDD '26]{Proceedings of the 32nd ACM SIGKDD Conference on Knowledge Discovery and Data Mining V.1}{August 09--13, 2026}{Jeju Island, Republic of Korea}
\acmBooktitle{Proceedings of the 32nd ACM SIGKDD Conference on Knowledge Discovery and Data Mining V.1 (KDD '26), August 09--13, 2026, Jeju Island, Republic of Korea}
\acmDOI{10.1145/3770854.3783958}
\acmISBN{979-8-4007-2258-5/2026/08}

\newcommand{\model}{MGOE}

\newcommand{\node}{macro nodes}
\newcommand{\edge}{macro edges}
\newcommand{\graphUpper}{MTMG}

\usepackage{amsmath}
\usepackage{bbm}
\usepackage{rotating}
\usepackage{multirow}
\usepackage{booktabs}
\usepackage{graphicx} 
\usepackage[para]{footmisc}
\usepackage[ruled, linesnumbered]{algorithm2e}

\usepackage{enumitem}
\usepackage{balance}
\usepackage{amsmath,amsfonts,bm}

\def\eqref#1{equation~\ref{#1}}
\def\1{\bm{1}}

\def\ve{{\bm{e}}}

\def\vz{{\bm{z}}}

\def\mE{{\mathbf{E}}}

\def\mK{{\mathbf{K}}}

\def\mM{{\mathbf{M}}}

\def\mQ{{\mathbf{Q}}}
\def\mR{{\mathbf{R}}}
\def\mS{{\mathbf{S}}}

\def\mV{{\mathbf{V}}}
\def\mW{{\mathbf{W}}}

\def\mY{{\mathbf{Y}}}
\def\mZ{{\mathbf{Z}}}

\DeclareMathAlphabet{\mathsfit}{\encodingdefault}{\sfdefault}{m}{sl}
\SetMathAlphabet{\mathsfit}{bold}{\encodingdefault}{\sfdefault}{bx}{n}

\def\gG{{\mathcal{G}}}

\newcommand{\sigmoid}{\sigma}

\begin{document}

\title{Macro Graph of Experts for Billion-Scale Multi-Task Recommendation}

\author{Hongyu Yao}
\authornote{Both authors contributed equally to this research.}
\affiliation{%
  \institution{Jinan University}
  \city{Guangzhou}
  \country{China}
}
\email{hongyuyao666@gmail.com}

\author{Zijin Hong} 
\authornotemark[1]
\affiliation{%
  \institution{The Hong Kong Polytechnic University}
  \city{Hong Kong SAR}
  \country{China}
}
\email{zijin.hong@connect.polyu.hk}

\author{Hao Chen}
\affiliation{%
  \institution{City University of Macau}
  \city{Macau SAR}
  \country{China}}
\email{sundaychenhao@gmail.com}

\author{Zhiqing Li}
\affiliation{%
  \institution{City University of Macau}
  \city{Macau SAR}
  \country{China}
}
\email{zhiqinglizzy@gmail.com}

\author{Qijie Shen}
\affiliation{%
  \institution{Alibaba Group}
  \city{Hangzhou}
  \country{China}}
\email{qjshenxdu@gmail.com}

\author{Zuobin Ying}
\affiliation{%
  \institution{City University of Macau}
  \city{Macau SAR}
  \country{China}
}
\email{zbying@cityu.edu.mo}

\author{Qihua Feng}
\affiliation{%
    \institution{Beijing Institute of Technology}
    \city{Beijing}
    \country{China}
}
\email{fengqh2020@gmail.com}

\author{Huan Gong}
\affiliation{%
  \institution{National University of Defense Technology}
  \city{Changsha}
  \country{China}
}
\email{gongh15@outlook.com}

\author{Feiran Huang}
\authornote{Corresponding author.}
\affiliation{%
  \institution{Beihang University}
  \city{Beijing}
  \country{China}
}
\affiliation{%
  \institution{Jinan University}
  \city{Guangzhou}
  \country{China}
}
\email{huangfr@ieee.org}

\renewcommand{\shortauthors}{Hongyu Yao et al.}

\begin{abstract}
Graph-based multi-task learning at billion-scale presents a significant challenge, as different tasks correspond to distinct billion-scale graphs. Traditional multi-task learning methods often neglect these graph structures, relying solely on individual user and item embeddings. However, disregarding graph structures overlooks substantial potential for improving performance. In this paper, we introduce the \textbf{Macro Graph of Experts (\model)} framework, the first approach capable of leveraging macro graph embeddings to capture task-specific macro features while modeling the correlations between task-specific experts. Specifically, we propose the concept of a \textbf{Macro Graph Bottom}, which, for the first time, enables multi-task learning models to incorporate graph information effectively. 
We design the Macro Prediction Tower to dynamically integrate macro knowledge across tasks.
\model \ has been deployed at scale, powering multi-task learning for a leading billion-scale recommender system, Alibaba. Extensive offline experiments conducted on three public benchmark datasets demonstrate its superiority over state-of-the-art multi-task learning methods, establishing \model \ as a breakthrough in multi-task graph-based recommendation. Furthermore, online A/B tests confirm the superiority of \model \ in billion-scale recommender systems.
\end{abstract}

\begin{CCSXML}
<ccs2012>
    <concept>
       <concept_id>10002951.10003317.10003347.10003350</concept_id>
       <concept_desc>Information systems~Recommender systems</concept_desc>
       <concept_significance>500</concept_significance>
    </concept>
    <concept>
       <concept_id>10002951.10003317.10003347.10011712</concept_id>
       <concept_desc>Information systems~Business intelligence</concept_desc>
       <concept_significance>500</concept_significance>
    </concept>
    <concept>
       <concept_id>10002951.10003260.10003272</concept_id>
       <concept_desc>Information systems~Online advertising</concept_desc>
       <concept_significance>300</concept_significance>
    </concept>
</ccs2012>
\end{CCSXML}

\ccsdesc[500]{Information systems~Recommender systems}
\ccsdesc[500]{Information systems~Business intelligence}
\ccsdesc[300]{Information systems~Online advertising}

\keywords{multi-task learning, graph neural networks, billion-scale recommender system, click-through rate prediction}

\maketitle

\section{Introduction}
\begin{figure}[tbp]
  \centering
  \includegraphics[width=\columnwidth]{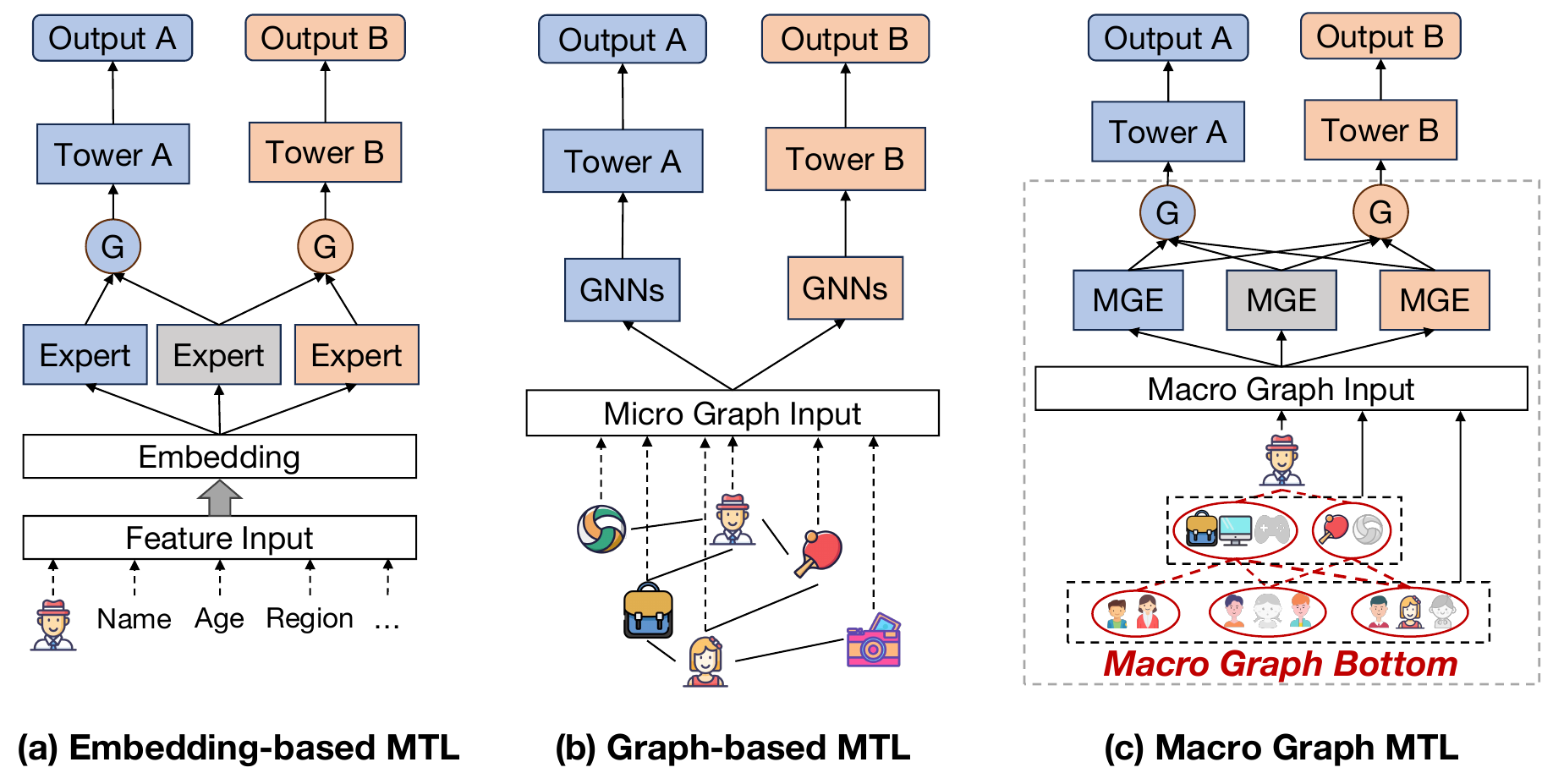} 
  \caption{Comparison of Multi-Task Learning Architectures, where "G" denotes a gating mechanism and "MGE" represents the proposed Macro Graph Expert.
(a) Embedding-based MTL relies on feature representations. 
(b) Graph-based MTL utilizes GNNs to capture relational semantics. 
(c) Macro Graph MTL introduces macro graph bottom to incorporate macro graph information.}
  \label{fig:intro}
\end{figure}

Multi-task learning (MTL) lies at the core of nearly all billion-scale recommendation systems~\cite{wang2023survey1, caruana1997multitask, kokkinos2017ubernet,zhang2023pogcn_workshop}, responsible for predicting multiple key tasks such as clicks, likes, and adding to cart~\cite{juan2016field, zhu2023final}. 
Its goal is to address multiple tasks in real-time to meet diverse user needs.
Previous multi-task recommendation models overlook the valuable information embedded within graphs. As shown in Figure~\ref{fig:intro}, their embedding-based multi-task bottoms rely solely on individual user and item embeddings and disregard the rich relational semantics encoded in graph structures, which ultimately results in suboptimal performance.
However, incorporating graph structures in MTL presents greater challenges compared to single-task situations due to two primary reasons: 1) online systems have billions of users and items, necessitating efficient and effective handling of this scale, and 2) multiple graph relations are formed by different tasks, requiring appropriate graph management techniques.

The mainstream approach in industrial recommendation typically employs single embeddings for multi-task problems without considering graph structures~\cite{sener2018multi, ruder2019latent,shen2025aliboost}.
ESMM~\cite{ma2018essm} enhances the robustness and improves the overall performance by utilizing the entire exposed sample space for training. 
MMoE~\cite{ma2018mmoe} pioneers the use of a basic mixture-of-experts structure, enabling task-specific predictions to learn from other tasks. 
PLE~\cite{tang2020ple} separates task-sharing and task-specific experts to avoid the negative transfer and seesaw phenomenon. 
STEM~\cite{su2024stem} introduces a novel shared and task-specific embeddings paradigm to effectively capture user preferences.
Subsequent works focused on designing more sophisticated expert structures, such as multi-layer or task-specialized experts. 
Other studies explore graph-based multi-task learning; for example, MoGENet~\cite{hu2022mogenet} proposes a multi-channel graph neural networks (GNNs) to model high-order information using user-item bipartite graphs. 
MMoCEG~\cite{zhang2023mmoceg} introduces graph-based MoE to recognize the commonalities and differences among multiple regions.
However, graph-based MTL models often struggle with the high computational burden of GNNs when facing billion-scale graphs.

In general, current multi-task learning models face trade-offs in billion-scale recommender systems:
\\
\textbf{1. Lack of High-Order Information}: Graph neural networks have demonstrated significant performance improvements, but single-embedding MTL approaches cannot benefit from extracting and aggregating high-order information.
\\
\textbf{2. Computational Complexity}: Since each task forms a different graph, managing multiple billion-scale graphs introduces considerable computational overhead for online recommendation systems.

Considering these trade-offs, designing a billion-scale graph neural network structure that addresses multi-task learning scenarios is a promising direction. Recently, MacGNN~\cite{chen2024macro} introduced a macro graph structure for single click-through rate (CTR) prediction tasks. However, applying macro structures to multi-task learning presents the following challenges:
\\
\textbf{1. Multi-task Graph Design}: Designing appropriate graph structures to capture and integrate information from multiple tasks is a complex challenge.
\\
\textbf{2. Macro Expert Design}: Developing experts capable of learning graph information to predict different tasks requires careful consideration and architectural innovation.
\\
\textbf{3. Inter-Macro Merging}: Efficiently merging information across macro structures is crucial for effective multi-task learning.

By addressing these challenges, we introduce \textbf{Macro Graph of Experts (\model)}, the first graph neural network architecture for billion-scale multi-task recommendation that avoids introducing unbearable computational complexity. We present the \textbf{Macro Task Merging Graph (MTMG)}, which extracts multi-task information using a single macro graph. We then describe the design of macro graph experts that learn graph information to predict different tasks. Finally, we introduce the Macro Prediction Tower, which aggregates multi-level information to simultaneously predict multiple tasks.

Our contributions can be summarized as follows:

\begin{itemize}[leftmargin=*]
\item We pioneer the use of graph neural networks for billion-scale multi-task recommendation, solving the computational complexity challenges of GNNs in real-world applications.
\item We present the design of \graphUpper \ that incorporate multiple different graphs, fundamentally addressing the inability of existing multi-task models to utilize graph information.
\item We provide implementation details for deploying \model \ in a real-world billion-scale recommender system.
\item Extensive offline experiments conducted on three public benchmark datasets verify that \model \ outperforms both embedding-based multi-task learning models and graph-based recommendation models. 
\footnote{Source code is available at \url{https://github.com/RainmannnnN/MGOE}}
Besides, online A/B test have confirmed the superiority of \model \ in the billion-scale recommender systems. 

\end{itemize}

\section{Preliminaries}

\subsection{Notations}
We use $\mathcal{U}$, $\mathcal{I}$, and $\mathcal{T}$ to denote the user set, item set, and task set, respectively.
Each task $t \in \mathcal{T}$ is associated with a task-specific interaction matrix $\mR^{(t)}$.
For any user $u \in \mathcal{U}$ and item $i \in \mathcal{I}$, the entry $r^{(t)}_{ui} \in \mR^{(t)}$ represents the interaction between $u$ and $i$ under task $t$.
Specifically, $r^{(t)}_{ui} = 1$ indicates that user $u$ has interacted with item $i$, while $r^{(t)}_{ui} = 0$ indicates no interaction.
Each task $t$ is also associated with a dataset $\mathcal{D}_t$, where each element is represented as $(u, i, y_{ui}^{(t)}) \in \mathcal{D}_t$.
Here, the label $y_{ui}^{(t)} \in \{0, 1\}$ indicates whether user $u$ interacted (1) or did not interact (0) with item $i$ in task $t$.
$r^{(t)}_{ui}$ and $y_{ui}^{(t)}$ are numerical equivalent but have representations.

\subsection{Multi-Task Learning}
Multi-task learning aims to develop the ability to handle multiple related but distinct recommendation tasks simultaneously, which may involve predicting whether a user will interact with an item by performing a specific behavior (e.g., clicking on a product, adding it to a cart, or making a purchase). 
While these tasks share foundational latent factors, including user preference structures and item attribute representations, each corresponds to distinct decision stages and motivations, exhibiting significant task-specific characteristics.
The target of MTL is to learn a functions set $\mathcal{F}$, where each function $f_t \in \mathcal{F}$ takes user and item features as input to predict the probability $\hat{y}_{ui}^{(t)} \in [0, 1]$ of potential interaction under the corresponding task, which is defined as $f_t(\cdot): (u, i) \rightarrow \hat{y}_{ui}^{(t)}$.
Typically, MTL employs the binary cross-entropy loss as the loss function for each task $t$:
\begin{equation}
    \mathcal{L}_t = -\sum_{(u, i, y_{ui}^{(t)}) \in \mathcal{D}_t} \left[ y_{ui}^{(t)} \cdot \log(\hat{y}_{ui}^{(t)}) + (1 - y_{ui}^{(t)}) \cdot \log(1 - \hat{y}_{ui}^{(t)}) \right].
\end{equation}
The overall loss function $\mathcal{L}_{\text{MTL}}$ is obtained by weighted sum:
\begin{equation}
\label{eq:overall_loss}
    \mathcal{L}_{\text{MTL}} = \sum_{t \in \mathcal{T}} \alpha_t \cdot \mathcal{L}_t,
\end{equation}
where parameter $\alpha_t$ denotes the relative importance of task $t$. 

\begin{figure*}[!t]\label{framework}
\centering
    \includegraphics[width=\linewidth]{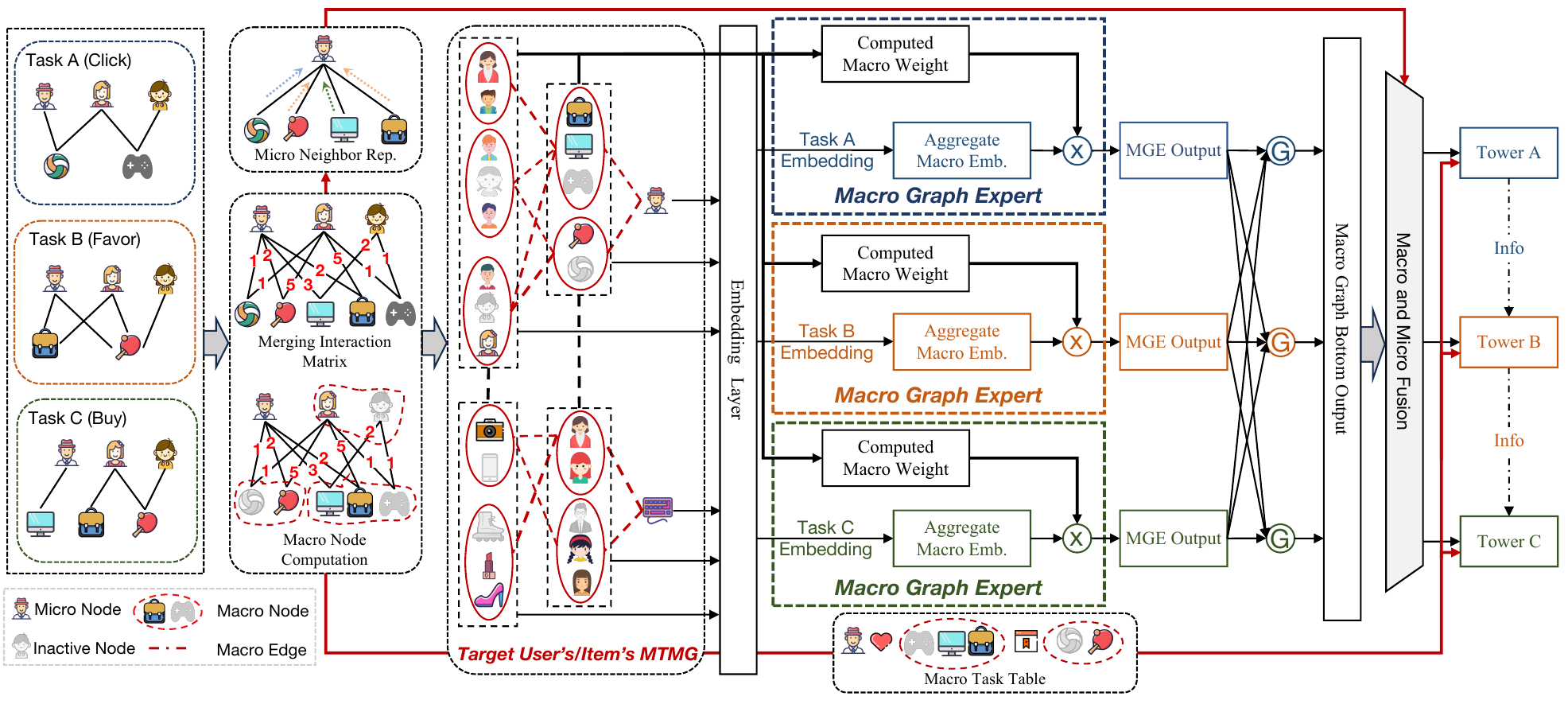}
    \caption{An illustration of the construction of Macro Task Merging Graph and the overview of Macro Graph of Experts.}
\label{fig:model_framework}
\end{figure*}

\subsection{MTL Bottom}
\paragraph{\textbf{Embedding-based bottom}} The embedding-based MTL bottom transforms input features into fused representations. 
For each task, the multi-task bottom takes vector $\vz = [\ve_{u};\;\ve_{i}]$ concatenated by relevant micro features from user $\ve_{u}$ and item $\ve_{i}$ as input, enabling further feature fusion for subsequent expert layers.
Suppose the number of shared experts and task-specific experts is $N_1$ and $N_2$, the predicted probability $\hat{y}_{ui}^{(t)}$ for task $t$ can be expressed by:
\begin{align}
h_t\left(\vz\right) &= \sum_{l=1}^{N_1} \operatorname{G}_{t,l}^{(1)} \left(\vz\right) \cdot \operatorname{E}_{s,l}\left(\vz\right)  + \sum_{l=1}^{N_2} \operatorname{G}_{t,l}^{(2)}\left(\vz\right) \cdot \operatorname{E}_{t,l}\left(\vz\right); \\ \label{eq:gate_function}
\hat{y}_{ui}^{(t)} &= \sigma\left(\operatorname{MLP}_t\left(h_t\left(\vz\right)\right)\right),
\end{align}
where $\sigmoid$ denotes the sigmoid function, 
$\operatorname{E}_{s,l}$ and $\operatorname{E}_{t,l}$ represent the $l$-th shared experts and task-specific experts for task $t$, respectively. $\operatorname{G}_{t,l}^{(1)}(\cdot)$ and $\operatorname{G}_{t,l}^{(2)}(\cdot)$ denote the gating network operations that fuse the outputs from the corresponding experts.
Note that the embedding-based MTL bottom focuses on micro features separately, neglecting the information induced by user-item interactions.

\paragraph{\textbf{Graph-based bottom}} 
The graph-based bottom captures the user-item collaborative connections through GNNs. 
Given the input of the $l$-th layer $\ve^l_v$ for a certain user/item node $v$, the output of $(l + 1)$-th layer $\ve^{l + 1}_v$ and the prediction $\hat{y}_{ui}^{(t)}$ for task $t$ are:  
\begin{align}
    \ve^{l + 1}_v &= \operatorname{GNN}\left(\ve^l_v, \sigmoid\left(\mW_{\gG_t} \cdot \sum_{c \in \mathcal{N}(v)} \eta_{v \leftarrow c} \cdot \ve^l_c \right)\right); \\
    \hat{y}_{ui}^{(t)} &= \sigmoid\left(\operatorname{MLP}_t\left([\ve^{l + 1}_u;\;\ve^{l + 1}_i]\right)\right),
\end{align}  
where $\mW_{\gG_t}$ is the weight matrix of the task graph $\gG_t$, $\eta_{v \leftarrow c}$ is implemented via attention, and $\mathcal{N}(v)$ denotes the set of neighbors of node $v$.
When applied to billion-scale task graphs, the graph-based bottom encounters significant computational challenges.

\subsection{Micro Graph}
A micro graph is a fine-grained structure that models user-item interactions individually, where users and items are treated as micro nodes. Micro edges are instantiated through actual interactions (e.g., purchases) and are represented by an interaction matrix.
Micro features characterize both micro nodes and edges. The trainable embeddings of micro nodes encode latent semantic information, while the entries of the interaction matrix $\mR$ quantify the properties of micro edges. This structure captures granular user-item relationships and serves as a foundational component in GNN-based recommendation models, enabling fine-grained interaction analysis and prediction.
Formally, a micro graph is denoted as $\mathcal{G} = (\mathcal{U}, \mathcal{I}, \mR)$.

\section{Methodology}
The overall framework of our proposed \model \ is illustrated in Figure~\ref{fig:model_framework}. In this section, we firstly introduce the utilized \graphUpper. 
Subsequently, we present the \model \ with detailed structure.

\subsection{Macro Task Merging Graph}
We begin by constructing \graphUpper, which provides essential macro-level graph information for the subsequent \model \ modeling.
The construction of \graphUpper \ primarily involves two core steps: (1) obtaining a merged interaction matrix, and (2) computing the macro nodes and constructing the macro edges in order.

\paragraph{\textbf{Merging Interaction Matrix}}
In real-world scenarios, multiple tasks often coexist (e.g., users may click on and purchase an item simultaneously).
These task combinations reflect the coexistence of diverse interaction patterns between users and items, indicating varying degrees of user preference.
To accurately capture these preferences, we introduce a scoring function to quantify the importance level of such task combinations.
For any user $u \in \mathcal{U}$ and item $i \in \mathcal{I}$, the potential mutual task combinations set is represented by:
\begin{equation}\nonumber
    \mathbf{C}_{ui} = \{\left(u, i, t_{j_{1}}\right), \left(u, i, t_{j_{2}}\right), \ldots, \left(u, i, t_{j_{k}}\right)  \mid t_{j_{1}}, t_{j_{2}}, \ldots ,t_{j_{k}} \in \mathcal{T}\}.
\end{equation}
Let the collection of $\mathbf{C}_{ui}$ as $\mathcal{C}=\{\mathbf{C}_{ui} \mid \ u\in\mathcal{U},i\in\mathcal{I}\}$. Once we have defined the importance order of each task, we can obtain a rank function for task combination sets~\cite{zhang2024pogcn}, denoted as $\rho(\cdot):\mathcal{C}\to\mathbb{N}^+$. We further define a scoring function $g(\cdot): \mathcal{C}\to \mathbb{R}$, which assigns a merging preference score to each task combination, representing its relative importance.
Specifically, the merging preference score $s_{ui}$ between user $u$ and item $i$ is defined as:
\begin{equation}
    \label{eq:scoring_function}
    s_{ui} = g\left(\mathbf{C}_{ui}\right) = 
    \begin{cases} 
\dfrac{[\rho(\mathbf{C}_{ui})+1]^{\beta}-1}{\beta} &, \text{if } \beta \neq 0 \\
\ln[\rho(\mathbf{C}_{ui})+1] &, \text{if }\beta=0
\end{cases}\,,
\end{equation}
where $\beta$ is a transformation parameter used to adjust the importance of $\mathbf{C}_{ui}$.
Then, we arrange these values corresponding to user-item pairs to construct the merging interaction matrix $\mS$:
\begin{equation}
    \mS = [s_{ui}]_{m \times n} ,
\end{equation}
where $s_{ui}$ denotes each entry of the matrix, and $m=\left|\mathcal{U}\right|$ and $n=\left|\mathcal{I}\right|$ represent the number of users and items, respectively.
The scoring function enables systematic and quantitative evaluation of task combinations, ensuring that \graphUpper\ accurately captures complex patterns in user-item interactions.
This enhances both the expressiveness of the macro graph and the efficiency of \model\ in downstream MTL.
Following the definition of the scoring function, we proceed to construct the nodes and edges within \graphUpper.

\paragraph{\textbf{Macro Node Computation}}
Computing a macro node involves leveraging the merging preference score computed by Equation~\ref{eq:scoring_function}.
Existing approaches treat all nodes in the macro graph uniformly, ignoring the rich semantic relationships in different task combinations.
To overcome this limitation, we propose Preference-Based Grouping, a novel method that assigns distinct weights to each micro node based on its merging preference score. Specifically, we construct \node\ from two complementary perspectives.

Let $\mS^\mathcal{U}=\mS\mS^\top$ and $\mS^\mathcal{I}=\mS^\top\mS$.
Then we can obtain macro nodes of by performing clustering on users and items based on $\mS^\mathcal{U}$ and $\mS^\mathcal{I}$, respectively. The following takes clustering of users as an example. Let $d_{vj}$ denote the $(v,j)$-th element of matrix $\mS^\mathcal{U}$.
Subsequently, we assign different weights to different micro nodes.
We define a merging preference score baseline $s_0$ for node $v$, such that if the sum of the merging preference score values exceeding $s_0$ surpasses a preset threshold  $\delta_{1}$, the node is considered active, and a relatively high weight $w^{+}$ is assigned. 
Conversely, if the sum falls below a certain value $\delta_{2}\,\left(\delta_{2} \leq \delta_{1}\right)$, the node is regarded as inactive, and a relatively low weight $w^{-}$ is assigned. 
For all other cases, a weight of 1 is assigned. 
Mathematically, the weight assigned to micro nodes $v$ is defined as:
\begin{equation}
    w_{v} = \begin{cases} 
    w^{+}\,, &\sum_{j=1}^{m} \mathbbm{1}_{\{d_{vj}\ge s_0\}}\geq \delta_{1} \\
    w^{-}\,, &\sum_{j=1}^{m} \mathbbm{1}_{\{d_{vj}\ge s_0\}} \le \delta_{2}  \\
    1, & \text{otherwise}
    \end{cases}\,,
    \label{eq:threshold}
\end{equation}
where $\mathbbm{1}_{\{\cdot\}}$ denotes the indicator function.
Then, the centroid update formula of preference-based grouping can be expressed as:
\begin{equation}
    \bm{\mu}_k = \frac{\sum_{v \in \mathcal{V}_k} w_{v} \cdot \mathbf{d}_{v}}{\sum_{v \in \mathcal{V}_k} w_{v}}, 
    \label{eq:macro_process}
\end{equation}
where $\bm{\mu}_k$ is the centroid of cluster $k$, $\mathcal{V}_k$ denotes the set of micro nodes assigned to cluster $k$, $w_{v}$ is the weight associated with node $v$, and $\mathbf{d}_v=(d_{v1},d_{v2}\cdots,d_{vm})^\top$ is the $v$-th row of $\mS^{\mathcal{U}}$. The clustering of items can be performed using the same method based on $\mS^{\mathcal{I}}$.
Preference-based grouping ensures that micro nodes with larger weights exert greater influence on the centroid's position, thereby making the centroid determination more flexible and better aligned with users' actual interests and task patterns.

\paragraph{\textbf{Macro Edge Construction}}
Within the \graphUpper, macro edges illustrate the interactions between pairs of \node \ within defined user or item subgraphs, signifying the task trends within these subgraphs. 
Unlike micro edges, which link predetermined user and item nodes, macro edges are designed to adaptively represent the varying strengths of connections between two \node, customized for each user or item. 
Specifically, we formalize the entire \node \ set in the \graphUpper \ as
$\widetilde{\mathcal{V}}_{\text{\graphUpper}} = \left\{ \mathcal{V}_{k_1}, \mathcal{V}_{k_2}, \ldots, \mathcal{V}_{k_{\widetilde{m}+\widetilde{n}}} \right\}$ with $\widetilde{m}$ and $\widetilde{n}$ denote the number of user clusters and item clusters, respectively. 
For any node $v$ and its \( j^{\text{th}} \)-hop neighbors, the weight of the macro edges is denoted by:
\begin{equation}
    \widetilde{\mS}_{v;\;\widetilde{p},\,\widetilde{q}}^{(j)} = \sum_{a \in {\mathcal{V}}_{v;\;\widetilde{p}}^{(j-1)}} \sum_{b \in {\mathcal{V}}_{v;\;\widetilde{q}}^{(j)}} s_{ab}\,,
    \label{eq:macro_edge_weight}
\end{equation}
where $\mathcal{V}_{v;\;\widetilde{p}}^{(j-1)}$ = $\mathcal{V}_{v;\;\widetilde{p}} \ \cap \ \mathcal{N}_v^{(j - 1)}$ denotes the macro nodes related to node $v$ within its $(j - 1)^{\text{th}}$-hop neighbors and $\mathcal{V}_{v;\;\widetilde{q}}^{(j)}$ = $\mathcal{V}_{v;\;\widetilde{q}} \ \cap \ \mathcal{N}_v^{(j)}$ represent the macro nodes related to node $v$ within its $j^{\text{th}}$-hop neighbors.
By merging different task combinations, \graphUpper \ provides a macro-level view of user-item interactions. This approach allows us to observe overarching trends and structures that might not be evident from individual interactions alone.

\paragraph{\textbf{Macro Task Merging Graph Definition}}
Our proposed \graphUpper \ can be formally defined as $\widetilde{\gG}_{\text{\graphUpper}} = \left(\widetilde {\mathcal{U}}, \widetilde {\mathcal{I}}, \widetilde {\mS}\right)$, where $\widetilde {\mathcal{U}}$ represents the set of macro user nodes and $\widetilde {\mathcal{I}}$ denotes set of the macro item nodes.
Each macro node is associated with a trainable macro embedding $\widetilde{\textbf{E}}$, and its macro neighbors are represented as $\widetilde{\mathcal{N}}$.
\graphUpper \ captures task patterns from a macro perspective and provides macro graph information for the subsequent macro graph bottom.

\subsection{Macro Graph Bottom} 
In this paper, we propose the macro graph bottom (MGB) with a mixture of macro graph experts (MGE) to exploit rich information from a macro perspective. 
In MGB, for each task $t$ and a certain node $v$, we assign task-specific macro embeddings to adequately capture the unique macro characteristics:
\begin{equation}
    \widetilde{\ve_v^t} = \operatorname{Lookup}\left(v, \widetilde{\mE}_t\right),
\end{equation}
where $\widetilde{\ve_v^t}$ represent the macro embeddings.
In addition to its own micro features, MGB applies the tailored MGEs to extract information from macro neighbors, achieving a more comprehensive understanding of the relationships and interactions between entities in \graphUpper \ from a macro perspective. 
For a given user-item pair from $\mathcal{D}_t$ in \graphUpper, the input of MGB is concatenated by its micro embedding and macro neighbors embedding:
\begin{equation}
    \vz_t = [\ve_u;\;\ve_i;\; \widetilde{\mE}_{t}(\widetilde{\mathcal{N}}_u);\;\widetilde{\mE}_{t}(\widetilde{\mathcal{N}}_i)],
\end{equation}
where $\ve_u$ and $\ve_i$ represents the original micro features and  $\widetilde{\mE}_{b}(\widetilde{\mathcal{N}}_v)$ denotes the corresponding macro embeddings of $v$'s macro neighbors, 
and the task $t$ output of MGB is illustrated as follow:
\begin{equation}
    \operatorname{MGB}\left(\mathcal{D}_t, \widetilde{\mE}, \widetilde{\mathcal{N}}\right) = \sum_{{t'} \in \mathcal{T}} \operatorname{G}_{t'}^{(t)}\left( \operatorname{MGE}_t(\vz_t) \right) \cdot \operatorname{MGE}_{t'}\left(\vz_{t'}\right),
\end{equation}
where $\operatorname{G}_{t'}^{(t)}(\cdot)$ denotes the gating network operation of task $t$, and $\operatorname{MGE}_{t'}(\cdot)$ is defined as Equation~\ref{eq:MGE}.
$\operatorname{MGB}$ integrates user and item embeddings with their corresponding macro neighbor embeddings, resulting in a more informative representation that reflects the overall macro structure.

\subsection{Macro Graph Expert}
Embedding-based multi-task experts, implemented using an MLP, face limitations in effectively leveraging the macro information in \graphUpper. 
We introduce a tailored MGE designed to exploit macro information efficiently. 
Specifically, we compute macro weights to determine the preference of the target node $v$ with respect to its macro neighbors. 
These macro weights are derived from the connected \edge. 
Furthermore, different task patterns can cause significant disparities in macro edge weights, which we address using logarithmic smoothing.
Formally, for a target node $v$  with its $j^{\text{th}}$-hop macro neighbor $\widetilde{p}$, we apply logarithmic smoothing with a parameter $\gamma$ to measure the importance of these macro neighbors. 
The macro weight $w_{v;\;\widetilde{p}}^{(j)}$ is computed as follows:
\begin{equation}
w_{v;\;\widetilde{p}}^{(j)} = \operatorname{softmax} \left( \log \left( \sum_{\widetilde{q} \in \widetilde{\mathcal{N}}_{v}^{(j-1)}} \widetilde{\mS}_{v;\;\widetilde{p},\,\widetilde{q}}^{(j)} + 1 \right), \gamma \right),
\label{eq:macro_weight}
\end{equation}
where $\gamma$ is a hyperparameter controlling the degree of smoothing~\cite{chen2023tau}. 
To aggregate macro embeddings efficiently, we employ an attention mechanism~\cite{vaswani2017attention} instead of computationally expensive graph convolutional neural networks. 
The aggregated macro embedding \(\widetilde{\mZ}_{v, \widetilde{p}}\) is given by:
\begin{align} \label{eq:agg_macro_embedding}
    \mM_{\text{sim}} &= \frac{(\mQ_v \cdot \widetilde{\mE}_{\widetilde{p}}) \cdot \left(\mK_v \cdot \mE_v\right)^{\top}}{\sum_{\widetilde{n}\in \widetilde{\mathcal{N}}_v^{(j)} \setminus \{\widetilde{p}\}} (\mQ_v \cdot \widetilde{\mE}_{\widetilde{n}}) \cdot \left(\mK_v \cdot \mE_v\right)^{\top}}; \\
    \widetilde{\mZ}_{v, \widetilde{p}} &= \operatorname{softmax}(\frac{\mM_{\text{sim}}}{\sqrt{d}}) \cdot \left(\mV_v \cdot \widetilde{\mE}_{\widetilde{p}}\right),
\end{align}
where $\mQ_{v}, \mK_{v}$ and $\mV_{v}$ are learnable query, key, and value matrices for node $v$, and $\mM_{\text{sim}}$ reflects the similarity between target node $v$ and \node \ $\widetilde{p}$. 
Finally, the fusion of macro neighbor representations of task $t$ is expressed as:
\begin{equation} \label{eq:MGE}
    \operatorname{MGE}_t(\vz_t) = \sum_{\widetilde{p} \in \widetilde{\mathcal{N}}_v^{(j)} }w_{v,\widetilde{p}} \cdot \widetilde{\mZ}_{v,\,\widetilde{p}}\,,
\end{equation}
where $\operatorname{MGE}_t(\vz_t)$ represents the final output of macro graph expert.

\subsection{Macro Prediction Tower}
\paragraph{\textbf{Macro and Micro Fusion}}
To capture the detailed local interactions provided by micro neighbors and the broader context offered by macro neighbors, we introduce the macro and micro fusion to aggregate information from both perspectives. 
Specifically, for a given target node $v$ with its numerous neighbors $\mathcal{N}_{v}$ related to task $t$, we sum its micro neighbors.
Then, the micro neighbors representation $\ve_{v;\;\text{nbr}}^t$ can be expressed as:
\begin{equation} \label{eq:micro_nbr_agg}
    \ve_{v;\;\text{nbr}}^t  = \sum_{c \in \mathcal{N}_{v}} \langle \ve^t_c, \ve_v^t \rangle \cdot \ve^t_c,
\end{equation}
where $\langle \ve^t_c, \ve_v^t \rangle$ indicates the weight between the target node and its micro neighbors.
Then, we mix the macro and micro informative representations through a multi-gate expert layer:
\begin{align}
    \ve_{\text{cat}}^t &= [\ve_u;\;\ve_i;\;\operatorname{MGB}(\vz_t);\;\ve_{u;\;\text{nbr}}^{t};\;\ve_{i;\;\text{nbr}}^{t} ];
    \label{eq:E_cat} \\
    x^t &= h_t\left(\ve_{\text{cat}}^t\right), \label{eq:expert_layer}
\end{align}
where $\ve_{\text{cat}}$ represents the concatenated embedding, $h_t(\cdot)$ is defined as Equation~\ref{eq:gate_function}, and $x^t$ denotes the informative fused output from both the macro and micro levels.

To exploit possible connections between tasks and fully utilize a user's macro preferences for sparse tasks, we replace the traditional multi-task tower with a macro correlated layer. Specifically, we use a macro cascading readout module to explore the relationships across different tasks and apply macro task adjustment to align the recommendations with the user's macro preferences.

\paragraph{\textbf{Macro Cascading Readout Module}}
Traditional multi-task recommendation models~\cite{ma2018mmoe, tang2020ple} obtain results in a parallel manner, which ignores the possible connections between different tasks. 
To address this, we introduce the macro cascading readout module that generates the final output in a cascaded fashion. 
Specifically, the output of the previous task is fed into the next prediction tower as additional input. 
Suppose $\hat{y}_{ui}^{(t_{j-1})}$ is the output of the previous task, then $\hat{y}_{ui}^{(t_j)}$ can be expressed as:
\begin{equation}
    \label{eq:casading_readout}
    \hat{y}_{ui}^{(t_j)} = 
    \begin{cases} 
        \sigma(\operatorname{MLP}(x^{t_1}))\,, &t_j = t_1\\
        \sigma(\operatorname{MLP}(x^{t_j}) + w_{t_{j-1}} \cdot \hat{y}_{ui}^{(t_{j-1})})\,, &t_j \in \mathcal{T} \setminus \{t_1\}\\
    \end{cases}\,,
\end{equation}
where $w_{t_{j-1}}$ is a trainable parameter that controls the strength of learning the correlation between different tasks.

\paragraph{\textbf{Macro Task Adjustment}}
With the goal of recommending more precisely to the user's individual preferences and enhancing the model's ability to predict sparse tasks, we propose macro task adjustment to guide the final prediction. 
First, based on \graphUpper, we construct a user macro task lookup table according to different task interactions. 
Specifically, we associate various user's task interactions with the corresponding \node.
For a certain user $u$, the final prediction $\hat{\mathrm{Y}}_{ui}$ can be expressed as:
\begin{equation}
    \hat{\mathrm{Y}}_{ui} = \mW_{u} \odot \mY_{\text{cat}} = \sigma( T_u ) \odot [\hat{y}_{ui}^{(t_1)};\;\hat{y}_{ui}^{(t_2)};\;\ldots]\,,
    \label{eq:final_output}
\end{equation}
where $T_u$ is the users' macro task lookup table, $\mW_{u}$ denotes the user-specific macro task adjustment weights. The prediction is computed via the Hadamard product (element-wise multiplication) $\odot$.

\subsection{Model Training}
We utilize binary cross-entropy as the loss function during the model's training. 
Specifically, with the loss function $\mathcal{L}_t$ for a certain task $t$, the objective of MTL $\mathcal{L}_{\text{MTL}}$ is given by Equation~\ref{eq:overall_loss}.
Then, the overall objective function for \model \ is formulated as:
\begin{equation}
    \mathcal{L}_{\text{MGOE}} = \mathcal{L}_{\text{MTL}} + \lambda_1 \cdot \|\mathbf{\theta}\|_1 + \lambda_2 \cdot \|\mathbf{\theta}\|_2^2\,,
    \label{eq:final_loss}
\end{equation}
where $\lambda_1 \cdot \|\mathbf{\theta}\|_1$ and $\lambda_2 \cdot \|\mathbf{\theta}\|_2^2$ represent the $\ell_1$ and $\ell_2$ regularization terms, respectively, included to prevent overfitting.

\section{Experiments}
\begin{table*}[!t]
  \centering
  \caption{Multi-task recommendation performance comparison results. The best and second-best results in each row are highlighted in \textbf{bold} font and \underline{underlined}, respectively.}
  \resizebox{\linewidth}{!}{
    \begin{tabular}{ccc|cccccc|cccc|cc}
    \toprule
    \multicolumn{3}{c|}{Models} & MMoE & PLE  & ESMM  & AITM & STEM & MoME & MoGENet & MMoCEG & POGCN & MacGNN & \model & \textit{Improv.(\%)}  \\

    \midrule
    \multicolumn{1}{c}{\multirow{8}[8]{*}{\rotatebox{90}{Taobao}}} 
    & \multicolumn{1}{c}{\multirow{2}[2]{*}{Click}} 
    & AUC      & 0.5580 & 0.5578 & 0.5577 & 0.5584 & 0.5561 & 0.5691 & 0.5784 & 0.5720 & 0.5841 & \underline{0.5891} & \textbf{0.6158} & \textbf{4.53\%} \\ &       
    & Logloss  & 0.3730 & 0.3723 & 0.3749 & 0.3922 & 0.3765 & 0.4415 & 0.4897 & 0.4134 & 0.3667 & \underline{0.3641} & \textbf{0.3609} & \textbf{0.88\%} \\

    \cmidrule{2-15}          
    & \multicolumn{1}{c}{\multirow{2}[2]{*}{Favor}} 
    & AUC      & 0.5672 & 0.5663 & 0.6034             & 0.5688 & 0.6192 & \underline{0.6481} & 0.6412 & 0.6250 & 0.5693 & 0.5967 & \textbf{0.8016} & \textbf{23.68\%} \\ &       
    & Logloss  & 0.1525 & 0.1519 & \underline{0.1496} & 0.1628 & 0.1580             & 0.1737 & 0.2211 & 0.1607 & 0.1547 & 0.1497 & \textbf{0.1301} & \textbf{13.03\%} \\
    
    \cmidrule{2-15}          
    & \multicolumn{1}{c}{\multirow{2}[2]{*}{Cart}} 
    & AUC      & 0.5459 & 0.5469 & 0.5660 & 0.5493 & 0.5666 & 0.5744 & 0.5823 & 0.5679 & 0.5714 & \underline{0.5837} & \textbf{0.6652} & \textbf{13.96\%}  \\ &       
    & Logloss  & 0.2470 & 0.2472 & 0.2471 & 0.2666 & 0.2534 & 0.2889 & 0.3333 & 0.2679 & 0.2458 & \underline{0.2424} & \textbf{0.2351} & \textbf{3.01\%}  \\
    
    \cmidrule{2-15}         
    & \multicolumn{1}{c}{\multirow{2}[2]{*}{Buy}} 
    & AUC      & 0.5560 & 0.5550 & 0.5549 & 0.5686 & 0.5676 & 0.5836 & 0.6286 & 0.5971 & 0.6142 & \underline{0.6927} & \textbf{0.7098} & \textbf{2.47\%}  \\ &       
    & Logloss  & 0.0966 & 0.0963 & 0.0961 & 0.1012 & 0.1038 & 0.1158 & 0.1356 & 0.1012 & 0.0972 & \underline{0.0930} & \textbf{0.0907} & \textbf{2.47\%}  \\

    \cmidrule{2-15}         
    & \multicolumn{1}{c}{\multirow{2}[2]{*}{Average}} 
    & AUC      & 0.5567 & 0.5565 & 0.5705 & 0.5612 & 0.5773 & 0.5937 & 0.6076 & 0.5905 & 0.5847 & \underline{0.6155} & \textbf{0.6981} & \textbf{13.42\%} \\ &       
    & Logloss  & 0.2172 & 0.2169 & 0.2169 & 0.2307 & 0.2229 & 0.2549 & 0.2949 & 0.2358 & 0.2161 & \underline{0.2120} & \textbf{0.2042} & \textbf{3.68\%}  \\

    \midrule
    \multicolumn{1}{c}{\multirow{8}[8]{*}{\rotatebox{90}{QB-video}}} 
    & \multicolumn{1}{c}{\multirow{2}[2]{*}{Click}} 
    & AUC      & 0.7891 & 0.7914 & 0.7934 & 0.7936 & 0.7910 & 0.8289 & 0.7786 & 0.8386 & 0.8402 & \underline{0.8415} & \textbf{0.8474} & \textbf{0.70\%} \\ &       
    & Logloss  & 0.4569 & 0.4560 & 0.4547 & 0.4524 & 0.4593 & 0.4337 & 0.4731 & 0.4199 & 0.4192 & \underline{0.4191} & \textbf{0.4047} & \textbf{3.44\%} \\
    
    \cmidrule{2-15}          
    & \multicolumn{1}{c}{\multirow{2}[2]{*}{Like}} 
    & AUC      & 0.6468 & 0.6328 & 0.7361 & 0.7547 & 0.7539 & 0.7250 & 0.5787 & \underline{0.7665} & 0.6673 & 0.6786 & \textbf{0.8533} & \textbf{11.32\%} \\ &       
    & Logloss  & 0.0403 & 0.0400 & \underline{0.0350} & 0.0358 & 0.0408 & 0.0368 & 0.0727 & 0.0357 & 0.0428 & 0.0396 & \textbf{0.0300} & \textbf{14.29\%} \\
    
    \cmidrule{2-15}          
    & \multicolumn{1}{c}{\multirow{2}[2]{*}{Share}} 
    & AUC      & 0.5125 & 0.5016 & 0.5726 & 0.5122 & \underline{0.6403} & 0.5385 & 0.5524 & 0.6110  & 0.5270 & 0.6188             & \textbf{0.6925} & \textbf{8.15\%} \\ &       
    & Logloss  & 0.0088 & 0.0086 & 0.0081 & 0.0167 & 0.0107             & 0.0081 & 0.0243 & 0.0133 & 0.0143 & \underline{0.0080} & \textbf{0.0078} & \textbf{2.50\%} \\
    
    \cmidrule{2-15}         
    & \multicolumn{1}{c}{\multirow{2}[2]{*}{Follow}} 
    & AUC      & 0.6128 & 0.6067 & 0.6606                & 0.5000 & 0.6792 & 0.6299 & 0.6277 & \underline{0.6915} & 0.6336 & 0.6504 & \textbf{0.7424} & \textbf{7.36\%} \\ &       
    & Logloss  & 0.0073 & 0.0076 & \underline{0.0070}    & 0.0131 & 0.0076             & 0.0072 & 0.0230 & 0.0080 & 0.0131 & 0.0071 & \textbf{0.0069}          & \textbf{1.43\%} \\

    \cmidrule{2-15}         
    & \multicolumn{1}{c}{\multirow{2}[2]{*}{Average}} 
    & AUC      & 0.6403 & 0.6331 & 0.6906 & 0.6401 & 0.7161 & 0.6805 & 0.6343 & \underline{0.7269} & 0.6670 & 0.6973             & \textbf{0.7839} & \textbf{7.84\%} \\ &       
    & Logloss  & 0.1283 & 0.1281 & 0.1262 & 0.1295 & 0.1296             & 0.1215  & 0.1483 & 0.1192 & 0.1224 & \underline{0.1185} & \textbf{0.1123} & \textbf{5.23\%}  \\

    \midrule
    \multicolumn{1}{c}{\multirow{8}[8]{*}{\rotatebox{90}{QK-article}}} 
    & \multicolumn{1}{c}{\multirow{2}[2]{*}{Read}} 
    & AUC      & 0.7089 & 0.7070 & 0.7124 & 0.7080 & 0.7249 & 0.7415 & 0.7732 & 0.7667 & 0.7387 & \underline{0.7911} & \textbf{0.8006} & \textbf{1.20\%} \\ &       
    & Logloss  & 0.1421 & 0.1514 & 0.1164 & 0.1284 & 0.1139 & 0.1077 & 0.1128 & 0.1174 & 0.1119 & \underline{0.0894} & \textbf{0.0879}             & \textbf{1.68\%} \\
    
    \cmidrule{2-15}          
    & \multicolumn{1}{c}{\multirow{2}[2]{*}{Like}} 
    & AUC      & 0.7969 & 0.7937 & 0.8118 & 0.7964 & 0.8165 & 0.8028 & 0.8036 & 0.8338 & 0.7959 & \underline{0.8701}  & \textbf{0.9327} & \textbf{7.19\%} \\ &       
    & Logloss  & 0.0865 & 0.0936 & 0.0721 & 0.0883 & 0.0719 & 0.0775 & 0.0815 & 0.0830 & 0.0795 & \underline{0.0658}  & \textbf{0.0519}          & \textbf{21.12\%} \\
    
    \cmidrule{2-15}          
    & \multicolumn{1}{c}{\multirow{2}[2]{*}{Favor}} 
    & AUC      & 0.8222 & 0.8228 & 0.8344 & 0.8226 & 0.8286 & 0.8315 & 0.8171 & 0.8516 & 0.7911 & \underline{0.8891} & \textbf{0.9294} & \textbf{4.53\%} \\ &       
    & Logloss  & 0.0393 & 0.0403 & 0.0323 & 0.0409 & 0.0329 & 0.0360 & 0.0389 & 0.0349 & 0.0392 & \underline{0.0309} & \textbf{0.0256}          & \textbf{17.15\%} \\

    \cmidrule{2-15}         
    & \multicolumn{1}{c}{\multirow{2}[2]{*}{Follow}} 
    & AUC      & 0.7329 & 0.7326 & 0.6664 & 0.7427 & 0.7345 & 0.7404 & 0.6381 & 0.7620 & 0.7054 & \underline{0.8075} & \textbf{0.8174} & \textbf{1.23\%} \\ &       
    & Logloss  & 0.0088 & 0.0087 & 0.0077 & 0.0089 & 0.0073 & 0.0077 & 0.0157 & 0.0076 & 0.0126 & \underline{0.0072} & \textbf{0.0070} & \textbf{2.78\%} \\

    \cmidrule{2-15}         
    & \multicolumn{1}{c}{\multirow{2}[2]{*}{Average}} 
    & AUC      & 0.7652 & 0.7640 & 0.7562 & 0.7674 & 0.7761 & 0.7790 & 0.7580 & 0.8035 & 0.7578 & \underline{0.8394} & \textbf{0.8700} & \textbf{3.65\%} \\ &       
    & Logloss  & 0.0692 & 0.0735 & 0.0571 & 0.0666 & 0.0565 & 0.0572 & 0.0622 & 0.0607 & 0.0608 & \underline{0.0483} & \textbf{0.0431} & \textbf{10.77\%} \\

    \bottomrule
    \end{tabular}%
    }
  \label{tab:main_result}%
\end{table*}%

In this section, we conduct comprehensive experiments on three real-world datasets, aiming to answer the following research questions.
\textbf{RQ1}: How does \model \ perform compared to current state-of-the-art multi-task models? \textbf{RQ2}: What are the effects of the different components of \model \ on the overall performance? \textbf{RQ3}: In what way do the hyper-parameters in \model \ affect the overall performance? \textbf{RQ4}: How efficient is the proposed \model? \textbf{RQ5}: How does \model \ perform on billion-scale recommendation platforms?

\subsection{Experimental Setup}
\subsubsection{Datasets}
We conduct extensive experiments on three real-world benchmark datasets: \textbf{Taobao}~\cite{zhu2018taobao}, \textbf{QB-video}~\cite{Tenrec} and
\textbf{QK-article}~\cite{Tenrec}. 
The detailed description and statistics of these datasets are illustrated in Appendix~\ref{app:data_appendix}.

\subsubsection{Compared Baselines}
To verify the effectiveness of \model, we compare with thirteen relevant representative models, which can be roughly divided into three categories.
(i) \textit{Traditional Multi-task Models}: \textbf{MMoE}, \textbf{PLE}, \textbf{ESMM}, \textbf{AITM}, \textbf{STEM}, \textbf{MoME}. (ii) \textit{Graph-based Methods}: \textbf{MoGENet}, \textbf{MMoCEG}, \textbf{POGCN}, \textbf{MacGNN}. (iii) \textit{Single-Task Learning Methods}: \textbf{DeepFM}, \textbf{DIN}, \textbf{DIEN}. We leave the details of these baseline models in Appendix~\ref{app:baseline_appendix}.

\subsubsection{Implementation Details}
All baselines and \model \ are implemented based on Pytorch. In the offline experiments, the embedding size fix to 10 and use Xavier~\cite{glorot2010xavier} for initialization for all models. For \model, we set the batch size to 1024 and use Adam~\cite{kingma2014adam} for optimization. The learning rate is searched from \{1e-2, 5e-2, 1e-3, 5e-3\} and the regularization term is searched from \{1e-4, 5e-5, 1e-5\}.

\subsubsection{Evaluation Metrics}
We adopt two widely used multi-task evaluation metrics, AUC~\cite{fawcett2006auc} and Logloss~\cite{chen2016logloss}, for a comprehensive evaluation.
Higher values of AUC indicate better prediction performance of the model, while the lower values of Logloss imply better prediction accuracy.
We run the experiments five times with different seeds with standard deviation to prevent extreme cases, the result is shown in Table~\ref{tab:main_result_five} in Appendix~\ref{app:main_result_five}. Additionally, the results of the performance of single-task learning (STL) models are included in Table~\ref{tab:STL_result} in Appendix~\ref{app:slt_result}.

\subsection{Main Results (RQ1)}
In this subsection, we compare our tailored \model \ with ten state-of-the-art baseline models on three experimental datasets. The comparison results on the AUC and Logloss metrics are reported in Table~\ref{tab:main_result}. From the result, we can have the following observations:

\textbf{\model \ can achieve significant improvements over state-of-the-art methods across all types of tasks on all experimental datasets.} From Table \ref{tab:main_result}, we observe that the customized \model \ attains the highest average AUC values and the lowest average Logloss values across all task types, outperforming both traditional multi-task models and graph-based methods. Specifically, in terms of the AUC metric, \model \ surpasses the best baseline with average AUC improvements of 13.42\% on Taobao, 7.84\% on QB-video, and 3.65\% on QK-article. Regarding the Logloss metric, \model \ achieves average effective gains of 3.68\%, 5.23\%, and 10.77\% respectively. These comparison results validate that \model \ can produce more reasonable multi-task recommendation outcomes.

\textbf{Introducing graph information can effectively improve the performance of the models.} Based on the results, we can observe that graph-based models, like MMoCEG and MacGNN, generally yield better results compared to traditional multi-task models. This implies that traditional multi-task models treat users and items in isolation, overlooking the relationship between them.
Conversely, integrating information from the user-item bipartite graph can mirror the more profound interactions between users and items. This is of great significance for enhancing the overall recommendation performance.

\textbf{Introducing graph information from the micro level can not fully utilize the task patterns.} By comparing the micro graph based methods and macro graph models in Table~\ref{tab:main_result}, it can be observed that macro graph models are capable of achieving relatively superior performance in multi-task recommendation compared to micro graph ones. This indicates that the traditional microscopic graph fails to effectively translate the diverse tasks of users towards items into enhanced model performance. 
Consequently, there is a need to integrate these user task patterns in a novel manner and consider these user and item nodes from a macroscopic perspective, which serves as the impetus for proposing \graphUpper.

\begin{table}[!t]
\centering 
\caption{Ablation study results between \model \ with its three variants on Taobao and QB-video.} 
\resizebox{\linewidth}{!}{
    \begin{tabular}{ll|cc}
    \hline
    \multicolumn{2}{c|}{Variant}                          & \multicolumn{1}{l}{Avg. AUC ($\uparrow$)} & \multicolumn{1}{l}{Avg. Logloss ($\downarrow$)} \\ \hline
    \multirow{3}[5]{*}{\rotatebox{90}{Taobao}} 
                            & \textbf{\model}             & \textbf{0.6936}±0.0040             & \textbf{0.2057}±0.0010                 \\ \cline{2-4} 
                            & \textit{w/o preference} & 0.6919±0.0048                      & 0.2060±0.0009                          \\
                            & \textit{w/o adjustment}            & 0.6836±0.0047                      & 0.2080±0.0031                          \\
                            & \textit{w/o cascading}       & 0.6927±0.0058                      & 0.2066±0.0009                          \\
    
                            \hline
                            
    \multirow{3}[5]{*}{\rotatebox{90}{QB-video}}
                            & \textbf{\model}             & \textbf{0.7715}±0.0114             & \textbf{0.1124}±0.0020                 \\ \cline{2-4} 
                            & \textit{w/o preference} & 0.7700±0.0201                      & 0.1127±0.0023                          \\
                            & \textit{w/o adjustment}            & 0.7513±0.0162                      & 0.1140±0.0040                          \\
                            & \textit{w/o cascading}       & 0.7640±0.0135                      & 0.1142±0.0033           \\  
    
                            \hline
    \end{tabular}
}
\label{tab:ablation}
\end{table}

\subsection{Ablation Study (RQ2)}
In order to verify the effectiveness of key designed components in \model, we conduct the ablation study and compare \model \ with its three variants: (1) \textbf{\textit{w/o preference}} removes the preference-based grouping module when constructing \graphUpper, which ignores the task combination. (2) \textbf{\textit{w/o adjustment}} excludes the macro task adjustment to overlook users' macro interest. (3) \textbf{\textit{w/o cascading}}, which the final output of \model \ is in a parallel fashion.

We conduct ablation studies on Taobao and QB-video datasets. The average of AUC and Logloss computed over all tasks is employed as the evaluation indicator. The results are presented in Table \ref{tab:ablation}. By analyzing this table, we arrive at the following observations:
\begin{itemize}[leftmargin=*]
    \item \textbf{Distinguishing the micro nodes helps form a better macro graph.} Ignoring the task relations between users and items decreases the performance of \model. This indicates \graphUpper \ is a more suitable macro graph for multi-task recommendation.
    \item \textbf{The macro prediction tower facilitates the macro information into more accurate recommendations.} When the macro task adjustment is removed, the performance deteriorates, as the \node \ preserve a wealth of users' preference information. Removing the macro cascading readout module leads to a decline in performance. This indicates the necessity of representing the transfer of macro knowledge across different tasks.
    
\end{itemize}

\begin{figure}[!t]
  \centering
  \includegraphics[width=\linewidth, trim=0cm 0cm 0cm 0cm,clip]{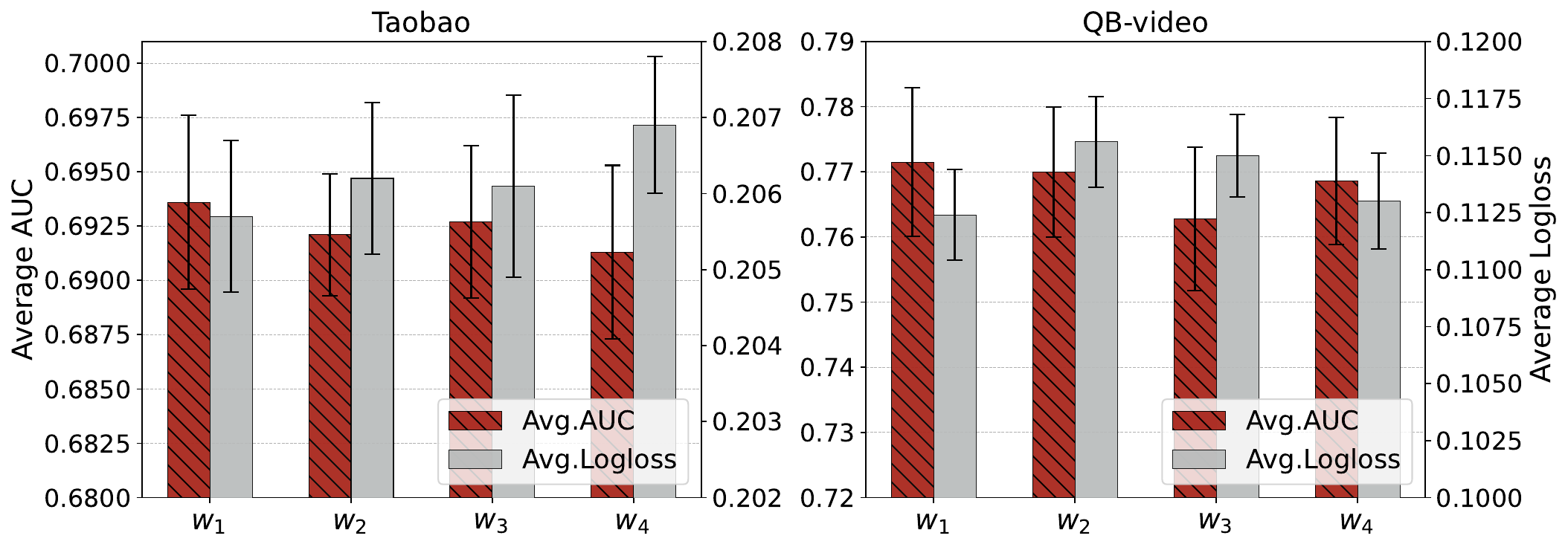}
  \caption{Parameter study of \(w_v\) on Taobao and QB-video.}
  \label{fig:param_w}
\end{figure} %

\begin{figure}[!h]
  \centering
  \includegraphics[width=\linewidth, trim=0cm 0cm 0cm 0cm,clip]{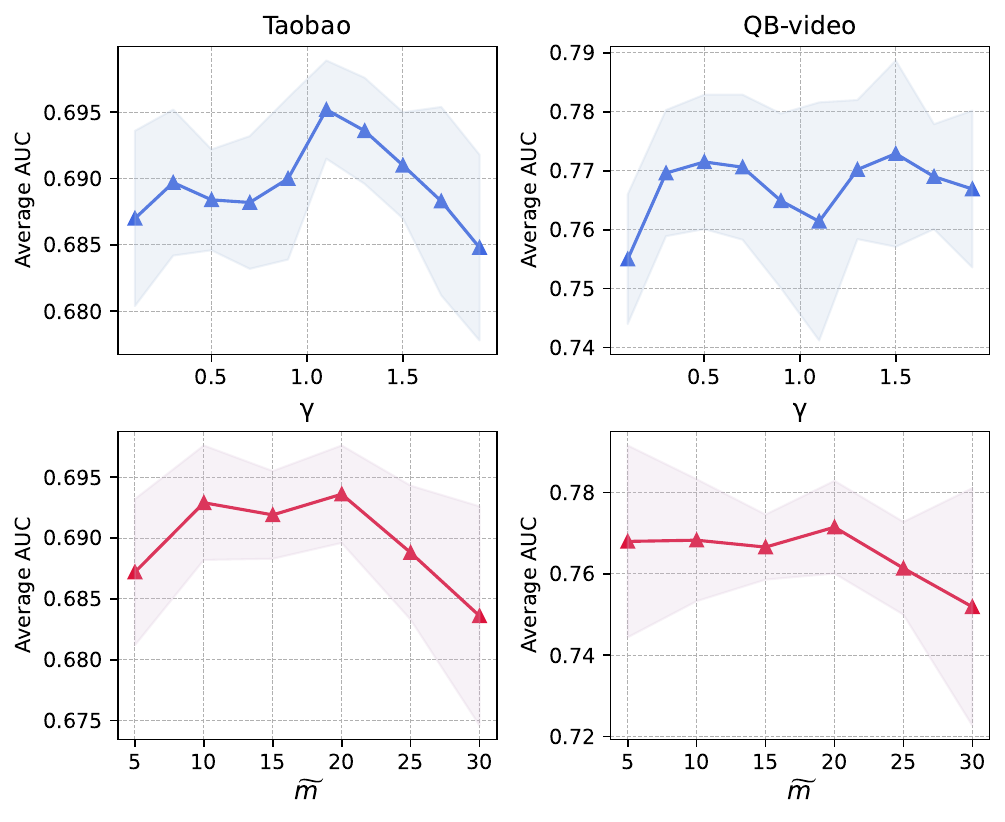}
  \caption{Parameter study of \(\gamma\) and \(\widetilde{m}\) on Taobao and QB-video.}
  \label{fig:param_tau_k}
\end{figure} %

\subsection{Hyper-parameter Study (RQ3)}
We investigate three key hyperparameters in \model: (1) the weight \(w_v\) in preference-based grouping; (2) the temperature parameter $\gamma$ for weighting \node; (3) the number of macro user nodes in \graphUpper, denoted as \(\widetilde{m}\). 
We conduct grid search within a small range over key hyperparameters on the Taobao and QB-video datasets. While examining $w_v, \gamma, \widetilde{m}$, we keep all other hyperparameters fixed. The results are shown in Figure ~\ref{fig:param_w} and Figure~\ref{fig:param_tau_k}.

\textbf{Effect of Preference-Based Grouping Weight}. 
We set the value of \(w^+\) to either 2 or 1, and \(w^-\) to either 1 or 0.5. Thus, we can represent the four combinations as follows: $w_1$=(2, 0.5), $w_2$=(2, 1),  $w_3$=(1, 0.5) and $w_4$=(1, 1), where the first value corresponds to $w^+$ and the second to $w^-$. From the results presented in Figure \ref{fig:param_w}, we can observe that in both datasets, assigning larger weights to active micro nodes and relatively smaller weights to inactive micro nodes generally leads to better performance.

\textbf{Effect of Temperature Parameter}.
We evaluate the impact of the temperature parameter $\gamma$ over the range from 0.1 to 1.9 with an increment of 0.2. The results are shown in the top half of Figure \ref{fig:param_tau_k}. From the line chart, we can observe that, in the Taobao dataset, as $\gamma$ increases, the model performance initially improves and then declines. The performance reaches its peak when \( \gamma=1.1 \).
In the QB-video dataset, \model \ performs best when \( \gamma=1.5 \), and then declines with the increase of \(\gamma\).

\textbf{Effect of Macro Node Number}.
We vary the number of macro user nodes from 5 to 30 in steps of 5.
From the results presented in the bottom half of Figure \ref{fig:param_tau_k}, we can observe that an overly small number of macro user nodes tends to lead to overly coarse segmentation, resulting in poor outcomes. Additionally, selecting a relatively appropriate number of \node \ can yield satisfactory performance.
For both Taobao and QB-video, when the number of macro user nodes reaches 20, \model \ achieves the best performance.

\begin{figure}[!t]
  \centering
  \includegraphics[width=\linewidth, trim=0cm 0cm 0cm 0cm,clip]{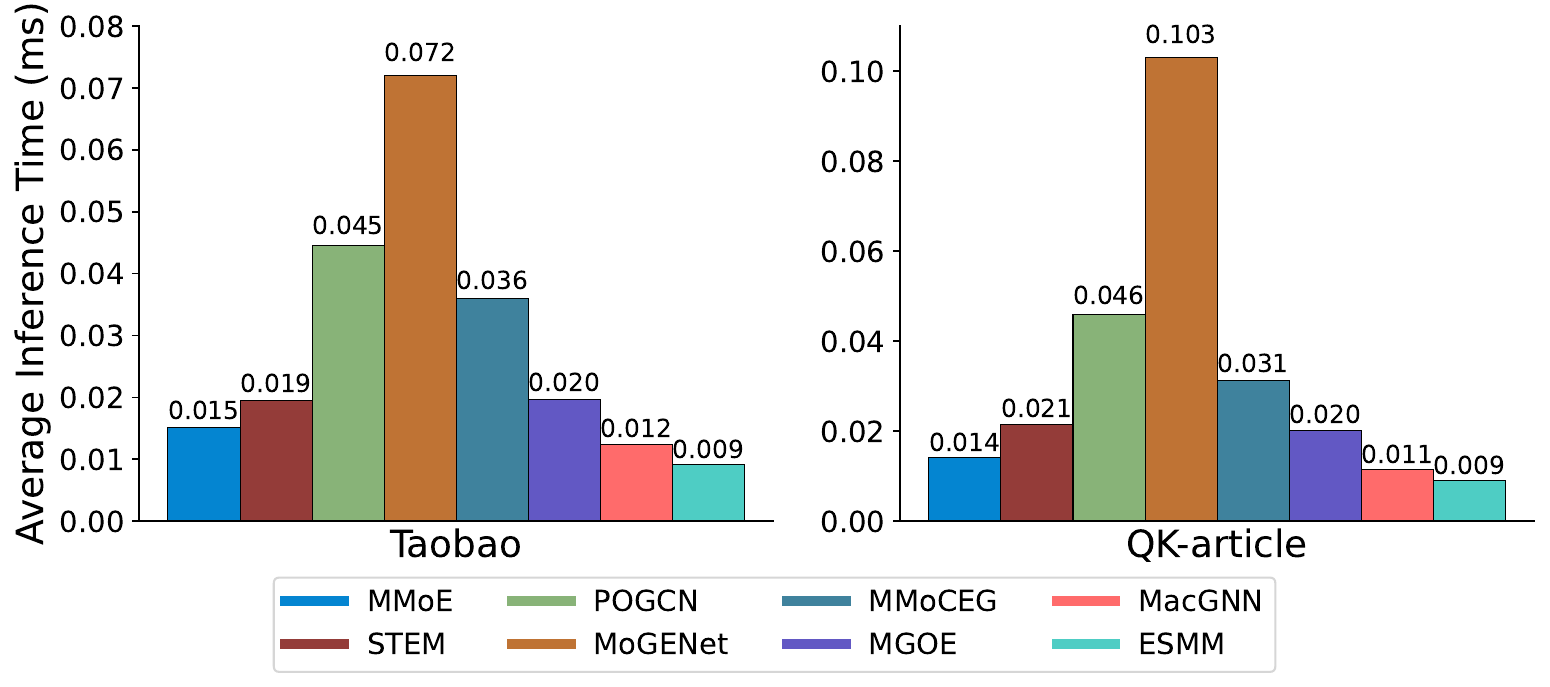}
  \caption{Efficiency study of the model inference time.}
  \label{fig:efficiency}
\end{figure} %

\subsection{Efficiency Study (RQ4)}
Since multi-task recommendation models are typically deployed in real-time scenarios, computational efficiency is a critical metric~\cite{wang2019learning}.
To verify the efficiency, we compare the average inference time, model parameter size, GPU memory usage, and training time between \model \ and several representative multi-task baselines.

\subsubsection{Average Inference Time}
The comparison results are presented in Figure~\ref{fig:efficiency}. From the figure, we have the following observations:
(i) \model \ is nearly as efficient as the simplest expert-based model and on par with STEM in terms of efficiency. While maintaining the same inference time as these models, our model incorporates high-order graph information, which significantly enhances the performance of multi-task recommendation.
(ii) \model \ has a much shorter inference time compared to graph-based models. Specifically, on the Taobao dataset, \model's inference time is 55.83\% faster than POGCN, 72.64\% faster than MoGENet and 45.13\% faster than MMoCEG. On the QK-article dataset, \model \ achieves an improvement of 56.31\%, 80.49\%, and 35.58\% respectively. This shows that our approach to integrating high-order graph information is much more computationally efficient, enabling faster and more responsive recommendations. 
(iii) Among all the models, those without expert networks (MacGNN, ESMM) have the fastest inference speed, as the main source of inference time comes from the expert architecture.

\begin{table}[!t]
    \centering
    \caption{Model parameter size, GPU memory usage and training time comparison on Taobao dataset.}
    \label{tab:params_gpu_time}
    \resizebox{\linewidth}{!}{
        \begin{tabular}{lccc}
            \toprule
            Models & Parameter Size/M & GPU Usage/MiB & Training Time/s \\
            \midrule
            MGOE     & 5.04  & 1194 & 258 \\
            MoGENet  & 11.57 & 2992 & 418 \\
            MMoCEG   & 12.22 & 1486 & 391 \\
            MMoE     & 3.21  & 1056 & 333 \\
            PLE      & 13.29 & 1302 & 957 \\
            STEM     & 13.59 & 1480 & 438 \\
            MacGNN   & 1.50  & 1080 & 320 \\
            \bottomrule
        \end{tabular}
    }
\end{table}

\subsubsection{Model Parameter Size}
In Table~\ref{tab:params_gpu_time}, we provide a detailed comparison of model parameter size. 
Among expert-based models, \model \ demonstrates advantages with a parameter size of 5.04M. Compared to other models such as PLE (13.29M) and MMoCEG (12.22M), \model \ has a significantly reduced parameter size, specifically reduced by approximately 62.08\% and 58.76\% respectively.
Although some models like MacGNN have a smaller parameter size (1.50M), \model \ maintain a balance between model complexity and performance, making it more suitable for real-world applications.

\subsubsection{GPU Memory Usage and Training Time}
Table~\ref{tab:params_gpu_time} presents the GPU memory usage (in MiB) and the training time (in seconds) of \model \ and other competing models on the Taobao dataset. 
Specifically, \model \ requires only 1194 MiB of GPU memory, 
which represents a reduction of approximately 60.10\% compared to MoGENet and 19.66\% compared to MMoCEG. Additionally, our model's GPU usage is also lower than that of PLE and STEM, and slightly higher than that of MMoE and MacGNN. These results demonstrate the efficiency of our model in terms of computational resource usage.
Besides, the training time of \model \ is remarkably shorter compared to other methods, indicating that MGOE can converge smoothly.

\subsection{Online Evaluation (RQ5)}
We conducted an online A/B test on Alibaba platform. In this experiment, our model served as a rank model, replacing the existing online best-performed ranking models---MMoE and MacGNN. MacGNN mainly considers the clicking task, MMoE and \model \ consider ``clicking'', ``adding to cart'', ``favoring'', and ``buying''. Table~\ref{tab:abtest} presents the average relative performance variation over two weeks for about 0.5 billion users and 1.2 billion items. 

\begin{table}[!t]
  \centering
  \caption{Results of online A/B tests in the Alibaba platform.}
  %\vspace{-0.5em}
  \resizebox{\linewidth}{!}{
    \begin{tabular}{l|ccccc}
    \toprule
    A/B Test & PCTR & UCTR & CVR & GMV & StayTime\\
    \midrule
    v.s.  MMoE &  +2.16\%   &  +1.63\%  & + 5.88\% & + 16.46\%  & +4.12\% \\
    v.s. MacGNN & +0.93\% & +0.75\% & +3.69\% & +7.74\% & +3.26\% \\
    \bottomrule
    \end{tabular}%
   }
  \label{tab:abtest}%
\end{table}%

\textit{Compared to MMoE}, firstly, \model \ demonstrates a performance improvement of 2.16\% for PCTR, 1.63\% for UCTR, 5.88\% for CVR, and 16.46\% for GMV, suggesting that our model boosts users’ inclination to interact with items and turn them into purchases. Secondly, the StayTime increases by 4.12\%. This indicates that, despite both models considering a wide range of user actions, by incorporating their comprehensive macro interests, our model is more effective in attracting user clicks and driving business revenue.

\textit{Compared to MacGNN}, which has a more narrow focus on clicking task, our model still shows favorable performance. The PCTR increased by 0.93\%, the UCTR rose by 0.75\%, the CVR increased by 3.69\%, the GMV grew by 7.74\%, and the user StayTime increased by 3.26\%. This suggests that the broader consideration of user actions in \model \ can lead to better overall business outcomes, even when compared to a model specialized in a single task.

Overall, in the online A/B test on the Alibaba platform, all the results show that \graphUpper \ and the equipped \model \ are more suitable than the state-of-the-art online multi-task recommendation models.

\section{Related Works}
\subsection{Multi-Task Learning}
Multi-task learning aims to enhance the model's generalization ability by training multiple tasks concurrently within the same model. 
In the early days of deep learning, the hard sharing mechanism of parameters was commonly employed. 
It could be applied to all hidden layers for all tasks, while maintaining task-specific output layers. 
The parameter soft sharing mechanism, on the other hand, allows each task to have its own model and parameters, thereby achieving better performance. 
To solve the issues of data sparsity (DS) and sample selection bias (SSB), ESSM~\cite{ma2018essm} simultaneously optimizes two related tasks in a sequential manner. MOE~\cite{MOE} employs a gated network to transform the original parameters shared by all samples into multiple sets of parameters, where each set is called an expert. MMOE~\cite{ma2018mmoe} proposes a multi-gate mixture-of-experts framework to explicitly learn to model task relationships.
PLE~\cite{tang2020ple} divides the model parameters into a private part and a shared part, which improves the robustness of multi-task learning and mitigates the negative transfer problem.

\subsection{Graph Learning for Recommendation}
In recent years, graphs have been increasingly incorporated into recommendation systems to enhance performance~\cite{ying2018graph,bei2025graphs,zhang2025cold,huang2025large}.
NGCF~\cite{wang2019ngcf} introduces higher-order connectivity into collaborative filtering, while LightGCN~\cite{he2020lightgcn} further improves efficiency by removing non-linear transformations.
However, the long inference time of such models limits their direct applicability to multi-task learning, motivating subsequent efforts to integrate graph structures into CTR prediction.
MoGENet~\cite{hu2022mogenet} adopts a multi-channel GNN to jointly model high-order signals on the user–item bipartite graph.
MMoCEG~\cite{zhang2023mmoceg} leverages a graph-based expert-sharing framework with contrastive learning to alleviate representation degeneration.
GLSM~\cite{sun2022glsm} employs adaptive fusion to combine long-term and short-term behavioral information.
GMT~\cite{min2022gmt} models heterogeneous local interactions to generate expressive embeddings for target user–item pairs.
MacGNN~\cite{chen2024macro} reduces computational cost by clustering similar micro-nodes into macro-nodes.

\section{Conclusion}
The introduction of the \textit{Macro Task Merging Graph (\graphUpper)} and \textit{Macro Graph of Experts (\model)} not only significantly advances the field of multi-task graph-based recommendation but also has practical implications for improving the performance of large-scale recommender systems. 
\graphUpper \ merges multiple billion-scale graphs into a unified macro structure, effectively consolidating complex graph data.
\model \ leverages macro graph embeddings with macro graph experts, which provides a novel approach for integrating macro graph information and enables more efficient information extraction.
Extensive offline experiments and online A/B tests demonstrate the superiority of our model.

\begin{acks}
This work was supported in part by the National Natural Science Foundation of China (Grant Nos. 62272200, 62502008, and 7250129), in part by the Funding Scheme for Research and Innovation of FDCT (Grant No. 0021/2025/ITP1), and in part by the Youth Independent Innovation Science Fund of the National University of Defense Technology (Grant No. ZK25-61).
\end{acks}

\bibliographystyle{acm}
\balance
\bibliography{reference}

\appendix

\section{Experimental Details}

\begin{table}[!h]
  \centering
  \caption{Statistics of the experimental datasets.}
  \resizebox{\linewidth}{!}{
    \begin{tabular}{c|c|c|c|c|c|c}
    \toprule
    Datasets & Users & Items & Click/Read & Cart/Like & Favor/Share & Buy/Follow \\
    \midrule
    Taobao & 28,910 & 65,087 & 1,391,096 & 103,524 & 55,141 & 29,909 \\
    QB-video & 28,803 & 15,540 & 1,503,207 & 14,261 & 1,955 & 1,675 \\
    QK-article & 52,052 & 30,273 & 1,560,571 & 28,073 & 11,275 & 1,811 \\
    \bottomrule
    \end{tabular}%
    }
  \label{tab:stats}%
\end{table}%

\begin{table*}[!h]
  \centering
  \caption{Multi-task recommendation performance comparison results over \textit{five} trial runs.}
  \resizebox{\linewidth}{!}{
    \begin{tabular}{ccc|cccccc|cccc|cc}
    \toprule
    \multicolumn{3}{c|}{Models} & MMoE & PLE  & ESMM  & AITM & STEM & MoME & MoGENet & MMoCEG & POGCN & MacGNN & \model & \textit{Improv.(\%)}  \\

    \midrule
    
    \multicolumn{1}{c}{\multirow{8}[8]{*}{\rotatebox{90}{Taobao}}} 
    & \multicolumn{1}{c}{\multirow{2}[2]{*}{Click}} 
    & AUC      & 0.5577±0.0005 & 0.5573±0.0006 & 0.5482±0.0026 & 0.5575±0.0002 & 0.5549±0.0009 & 0.5719±0.0040 & 0.5793±0.0138 & 0.5725±0.0049 & 0.5677±0.0049 & \underline{0.6039}±0.0009 & \textbf{0.6125}±0.0035 & \textbf{1.42\%} \\ &       
    & Logloss  & 0.3744±0.0017 & 0.3876±0.0016 &  0.3715±0.0003 & 0.3781±0.0032 & 0.3930±0.0105 & 0.4376±0.0176 & 0.4245±0.0273 & 0.4475±0.0791 & 0.3688±0.0004 & \underline{0.3678}±0.0011 & \textbf{0.3632}±0.0015 & \textbf{1.25\%} \\

    \cmidrule{2-15}          
    & \multicolumn{1}{c}{\multirow{2}[2]{*}{Favor}} 
    & AUC      & 0.5665±0.0013 & 0.6140±0.0011 & 0.5760±0.0099             & 0.5668±0.0011 & 0.6126±0.0026 & \underline{0.6493}±0.0106 & 0.6057±0.0365 & 0.6178±0.0125 & 0.5485±0.0108 & 0.6282±0.0048 & \textbf{0.7961}±0.0033 & \textbf{22.61\%} \\ &       
    & Logloss  & 0.1527±0.0006 & 0.1577±0.0025 & \underline{0.1493}±0.0004 & 0.1548±0.0016 & 0.1614±0.0036             & 0.1695±0.0074 & 0.1771±0.0094 & 0.1712±0.0239 & 0.1575±0.0038 & 0.1494±0.0008 & \textbf{0.1321}±0.0015 & \textbf{11.52\%} \\
    
    \cmidrule{2-15}          
    & \multicolumn{1}{c}{\multirow{2}[2]{*}{Cart}} 
    & AUC      & 0.5460±0.0008 & 0.5692±0.0010 & 0.5408±0.0025 & 0.5451±0.0008 & 0.5668±0.0016 & 0.5745±0.0047 & 0.5724±0.0177 & 0.5603±0.0044 & 0.5549±0.0031 & \underline{0.5935}±0.0026 & \textbf{0.6647}±0.0014 & \textbf{12.00\%}  \\ &       
    & Logloss  & 0.2479±0.0007 & 0.2589±0.0018 & 0.2456±0.0005 & 0.2533±0.0028 & 0.2627±0.0072 & 0.2899±0.0054 & 0.2949±0.0153 & 0.2962±0.0501 & 0.2485±0.0005 & \underline{0.2432}±0.0007 & \textbf{0.2359}±0.0004 & \textbf{3.00\%}  \\
    
    \cmidrule{2-15}         
    & \multicolumn{1}{c}{\multirow{2}[2]{*}{Buy}} 
    & AUC      & 0.5542±0.0014 & 0.5662±0.0041 & 0.5143±0.0027 & 0.5576±0.0039 & 0.5686±0.0010 & 0.5795±0.0061 & 0.6414±0.0135 & 0.5989±0.0045 & 0.6057±0.0033 & \underline{0.6869}±0.0069 & \textbf{0.7012}±0.0077 & \textbf{2.08\%}  \\ &       
    & Logloss  & 0.0968±0.0001 & 0.0993±0.0017 & 0.0970±0.0002 & 0.0995±0.0014 & 0.1096±0.0039 & 0.1132±0.0042 & 0.1140±0.0047 & 0.1075±0.0104 & 0.1020±0.0018 & \underline{0.0922}±0.0002 & \textbf{0.0915}±0.0005 & \textbf{0.76\%}  \\

    \cmidrule{2-15}         
    & \multicolumn{1}{c}{\multirow{2}[2]{*}{Average}} 
    & AUC      & 0.5561±0.0005 & 0.5767±0.0014 & 0.5448±0.0026 & 0.5567±0.0011 & 0.5757±0.0005 & 0.5938±0.0063 & 0.5997±0.0203 & 0.5874±0.0066 & 0.5692±0.0055 & \underline{0.6281}±0.0038 & \textbf{0.6936}±0.0040 & \textbf{10.43\%} \\ &       
    & Logloss  & 0.2179±0.0008 & 0.2259±0.0018 & 0.2159±0.0002 & 0.2214±0.0021 & 0.2317±0.0054 & 0.2526±0.0086 & 0.2526±0.0141 & 0.2556±0.0409 & 0.2192±0.0016 & \underline{0.2119}±0.0007 & \textbf{0.2057}±0.0010 & \textbf{2.93\%}  \\

    \midrule

    \multicolumn{1}{c}{\multirow{8}[8]{*}{\rotatebox{90}{QB-video}}} 
    & \multicolumn{1}{c}{\multirow{2}[2]{*}{Click}} 
    & AUC      & 0.7876±0.0031 & 0.7935±0.0009 & 0.7949±0.0002 & 0.7949±0.0003 & 0.7925±0.0008 & 0.8240±0.0044 & 0.6425±0.1038 & 0.8219±0.0220 & \underline{0.8350}±0.0035 & 0.8323±0.0040 & \textbf{0.8497}±0.0070 & \textbf{1.76\%} \\ &       
    & Logloss  & 0.4592±0.0021 &  0.4564±0.0008 & 0.4537±0.0003 & 0.4549±0.0011 & 0.4577±0.0032 & 0.4400±0.0045 & \text{3.7794±3.9909} & 0.4348±0.0176 & \underline{0.4222}±0.0035 & 0.4402±0.0105 & \textbf{0.4037}±0.0067 & \textbf{4.38\%} \\
    
    \cmidrule{2-15}          
    & \multicolumn{1}{c}{\multirow{2}[2]{*}{Like}} 
    & AUC      &  0.6385±0.0037 & 0.6279±0.0039 & 0.6251±0.0075 & 0.7518±0.0113 & \underline{0.7593}±0.0054 & 0.7396±0.0112 & 0.5503±0.0343 & 0.7396±0.0302 & 0.7264±0.0076 & 0.6953±0.0143 & \textbf{0.8431}±0.0090 & \textbf{11.04\%} \\ &       
    & Logloss  & 0.0406±0.0001 & 0.0380±0.0010 & 0.0367±0.0001 & 0.0361±0.0010 & 0.0388±0.0022 & \underline{0.0342}±0.0002 & 0.1509±0.0810 & 0.0359±0.0016 & 0.0379±0.0005 & 0.0404±0.0008 & \textbf{0.0310}±0.0012 & \textbf{9.36\%} \\
    
    \cmidrule{2-15}          
    & \multicolumn{1}{c}{\multirow{2}[2]{*}{Share}} 
    & AUC      & 0.5105±0.0041 &  0.5296±0.0180 & 0.5099±0.0017 & 0.5523±0.0080 & 0.5281±0.0562 & 0.5529±0.0345 & 0.5444±0.0095 & 0.5809±0.0382  & 0.5347±0.0075 & \underline{0.5866}±0.0230             & \textbf{0.6630}±0.0189 & \textbf{13.02\%} \\ &       
    & Logloss  & 0.0088±0.0001 & 0.0083±0.0002 & 0.0086±0.0001 & 0.0084±0.0001 & 0.0156±0.0023             & 0.0082±0.0001 & 0.0327±0.0055 & 0.0196±0.0140 & 0.0112±0.0005 & \underline{0.0081}±0.0001 & \textbf{0.0080}±0.0001 & \textbf{1.23\%} \\
    
    \cmidrule{2-15}         
    & \multicolumn{1}{c}{\multirow{2}[2]{*}{Follow}} 
    & AUC      & 0.6037±0.0035 & 0.6057±0.0069 & 0.5470±0.0070                & 0.6137±0.0035 & 0.6031±0.0031 & 0.5932±0.0368 & 0.5796±0.0476 & 0.6579±0.0200 & \underline{0.6678}±0.0053  & 0.6560±0.0211 & \textbf{0.7301}±0.0107 & \textbf{9.33\%} \\ &       
    & \text{Logloss}  & 0.0068±0.0001 & 0.0066±0.0001 & 0.0067±0.0002    & \underline{0.0063}±0.0001 & 0.0065±0.0001             & 0.0073±0.0002 & 0.0315±0.0062 & 0.0121±0.0093 & 0.0100±0.0004  & 0.0072±0.0001 & \textbf{0.0060}±0.0001          & \textbf{4.76\%} \\

    \cmidrule{2-15}         
    & \multicolumn{1}{c}{\multirow{2}[2]{*}{Average}} 
    & AUC      & 0.6351±0.0013 &  0.6392±0.0054 & 0.6145±0.0038 & 0.6781±0.0057 & 0.6707±0.0361 & 0.6774±0.0217 & 0.5792±0.0488 & \underline{0.7001}±0.0276 & 0.6910±0.0060 & 0.6926±0.0156             & \textbf{0.7715}±0.0114 & \textbf{10.20\%} \\ &       
    & Logloss  & 0.1289±0.0006 & 0.1273±0.0004 & 0.1264±0.0001 & 0.1264±0.0005 & 0.1296±0.0019             & 0.1224±0.0012  & 0.9986±1.0209 & 0.1256±0.0106 & \underline{0.1203}±0.0012 & 0.1240±0.0028 & \textbf{0.1124}±0.0020 & \textbf{6.57\%}  \\

    \midrule

    \multicolumn{1}{c}{\multirow{8}[8]{*}{\rotatebox{90}{QK-article}}} 
    & \multicolumn{1}{c}{\multirow{2}[2]{*}{Read}} 
    & AUC      & 0.6996±0.0008 &  0.7080±0.0020 & 0.7135±0.0034 & 0.7105±0.0047 & 0.7256±0.0006 & 0.7451±0.0045 & 0.7613±0.0073 & 0.7589±0.0019 & 0.7352±0.0028 & \underline{0.7741}±0.0049 & \textbf{0.7937}±0.0078 & \textbf{2.53\%} \\ &       
    & Logloss  & 0.1185±0.0013 & 0.1487±0.0028 & 0.1161±0.0005 & 0.1291±0.0032 & 0.1186±0.0029 & 0.1287±0.0081 & 0.1309±0.0182 & 0.1149±0.0037 & 0.1145±0.0024 & \underline{0.1047}±0.0088 & \textbf{0.0921}±0.0060             & \textbf{12.03\%} \\
    
    \cmidrule{2-15}          
    & \multicolumn{1}{c}{\multirow{2}[2]{*}{Like}} 
    & AUC      & 0.7768±0.0016 & 0.7947±0.0022 & 0.8107±0.0059 & 0.8003±0.0042 & 0.8184±0.0013 & 0.7892±0.0085 & 0.7322±0.0276 & 0.8284±0.0067 & 0.7763±0.0267 & \underline{0.8670}±0.0039  & \textbf{0.9250}±0.0074 & \textbf{6.69\%} \\ &       
    & Logloss  & 0.0782±0.0005 & 0.0933±0.0025 & 0.0722±0.0003 & 0.0845±0.0025 & 0.0741±0.0011 & 0.0776±0.0045 & 0.0905±0.0053 & 0.0770±0.0043 & 0.0811±0.0021 & \underline{0.0698}±0.0033  & \textbf{0.0532}±0.0023         & \textbf{23.78\%} \\
    
    \cmidrule{2-15}          
    & \multicolumn{1}{c}{\multirow{2}[2]{*}{Favor}} 
    & AUC      & 0.8001±0.0015 & 0.8231±0.0043 & 0.8303±0.0096 & 0.8227±0.0038 & 0.8287±0.0026 & 0.7996±0.0115 & 0.7447±0.0310 & 0.8480±0.0061 & 0.7760±0.0264 & \underline{0.8835}±0.0047 & \textbf{0.9351}±0.0070 & \textbf{5.84\%} \\ &       
    & Logloss  & 0.0369±0.0001 & 0.0402±0.0007 & \underline{0.0325}±0.0002 & 0.0395±0.0012 & 0.0341±0.0008 & 0.0343±0.0005 & 0.0468±0.0038 & 0.0354±0.0015 & 0.0409±0.0021 & 0.0342±0.0018 & \textbf{0.0251}±0.0004          & \textbf{22.77\%} \\

    \cmidrule{2-15}         
    & \multicolumn{1}{c}{\multirow{2}[2]{*}{Follow}} 
    & AUC      & 0.7040±0.0050 & 0.7385±0.0028 & 0.6936±0.0151 & 0.7421±0.0044 & 0.7304±0.0059 & 0.7150±0.0100 & 0.6295±0.0524 & 0.7597±0.0093 & 0.7072±0.0165 & \underline{0.7863}±0.0044 & \textbf{0.8225}±0.0085 & \textbf{4.60\%} \\ &       
    & Logloss  & 0.0078±0.0001 & 0.0088±0.0001 & 0.0076±0.0001 & 0.0089±0.0002 & \underline{0.0074}±0.0001 & 0.0076±0.0002 & 0.0193±0.0056 & 0.0077±0.0002 & 0.0154±0.0027 & 0.0075±0.0002 & \textbf{0.0066}±0.0001 & \textbf{10.81\%} \\

    \cmidrule{2-15}         
    & \multicolumn{1}{c}{\multirow{2}[2]{*}{Average}} 
    & AUC      & 0.7451±0.0014 & 0.7661±0.0015 & 0.7620±0.0024 & 0.7689±0.0038 & 0.7758±0.0014 & 0.7622±0.0086 & 0.7169±0.0296 & 0.7988±0.0060 & 0.7487±0.0181 & \underline{0.8277}±0.0045 & \textbf{0.8691}±0.0070 & \textbf{5.00\%} \\ &       
    & Logloss  & 0.0603±0.0005 & 0.0728±0.0011 & 0.0571±0.0002 & 0.0655±0.0002 & 0.0586±0.0011 & 0.0620±0.0033 & 0.0719±0.0082 & 0.0588±0.0024 & 0.0630±0.0023 & \underline{0.0540}±0.0035 & \textbf{0.0442}±0.0022 & \textbf{18.15\%} \\

    \bottomrule
    \end{tabular}%
    }
  \label{tab:main_result_five}%
\end{table*}%

\subsection{Dataset Details}\label{app:data_appendix}
We adopt three publicly available datasets for offline evaluation. 
The statistics of these datasets are presented in Table~\ref{tab:stats}. 
The detailed descriptions of the datasets are as follows:
\begin{itemize}[leftmargin=*]
    \item \textbf{\textit{Taobao}}\footnote{\url{https://tianchi.aliyun.com/dataset/649}}~\cite{zhu2018taobao} is a user behavior dataset provided by Alibaba, one of the most famous online shopping malls in China. It includes the activities between 28,910 users and 65,087 items from November to December 2017, including clicks, favors, carts, and buys. 

    \item \textbf{\textit{QB-video}}\footnote{\url{https://github.com/yuangh-x/2022-NIPS-Tenrec/tree/master}}~\cite{Tenrec} is a large-scale real-world dataset obtained from the recommendation data of two content platform applications of Tencent. 
    We select a subset of video scenes that consists of 28,803 users and 15,540 items, including four basic behaviors: click, like, share, and follow. 
    
    \item \textbf{\textit{QK-article}}\footnotemark[3]~\cite{Tenrec} is a content dataset under the article scene provided by Tencent. It records the behaviors of 52,052 users and 30,273 products, which mainly fall into four categories: reading, liking, favoring, and following.
    
\end{itemize}

\subsection{Baseline Details}\label{app:baseline_appendix}
We compare our proposed \model \ with thirteen relevant representative models, which can be roughly divided into three categories. 

\textbf{Traditional Multi-Task Models}: (i) \textbf{\textit{MMoE}}~\cite{ma2018mmoe} introduces a gating network to assign different expert networks to different tasks. (ii) \textbf{\textit{PLE}}~\cite{tang2020ple} extracts deep information layer by separating shared experts and task specific experts. (iii) \textbf{\textit{ESMM}}~\cite{ma2018essm} solves the problem of data sparsity and sample selection bias and improves generalizability by training the entire sample space. (iv) \textbf{\textit{AITM}}~\cite{xi2021aitm} proposes an adaptive information transfer framework to model sequence dependencies. (v) \textbf{\textit{STEM}}~\cite{su2024stem} proposes a share and task-specific embedding paradigm 
to facilitate the learning of task-specific embeddings and direct knowledge transfer across tasks. (vi) \textbf{\textit{MoME}}~\cite{xu2024mome} designs an expert to share parameters efficiently and maximize resource utilization.

\textbf{Graph-based Methods}: (i) \textbf{\textit{MoGENet}}~\cite{hu2022mogenet} introduces a multi-channel GNN module to extract information from the bipartite graph. (ii) \textbf{\textit{MMoCEG}}~\cite{zhang2023mmoceg} introduces a graph-based gating mechanism to recognize commonalities and differences between multiple regions. (iii) \textbf{\textit{POGCN}}~\cite{zhang2024pogcn} introduces partial order graphs utilizing GNN to consider behavior relations. (iv) \textbf{\textit{MacGNN}}~\cite{chen2024macro} solves computational complexity problems in the infrastructure by reducing the number of nodes from billions to hundreds.

\textbf{Single-Task Learning Methods}: (i) \textbf{\textit{DeepFM}}~\cite{guo2017deepfm} combines a factorization machine and a DNN to model user-item relationships. (ii) \textbf{\textit{DIN}}~\cite{zhou2018DIN} introduces an attention mechanism to capture the correlation between user's historical behavior and candidate ads. (iii) \textbf{\textit{DIEN}}~\cite{Zhou2019DIEN} further improves the accuracy of the CTR prediction by capturing the evolution of the user interests.

\subsection{Experimental Results with Five Runs}\label{app:main_result_five}
This subsection provides detailed results of experiments where all models were run five times independently on each dataset. The reported metrics are presented as mean ± standard deviation, ensuring statistical robustness and reliability of the performance comparisons. 
From Table \ref{tab:main_result_five}, we observe that \model \ also attains the highest average AUC values and the lowest average Logloss values across all task types. Specifically, in terms of the AUC metric, \model \ surpasses the best baseline with average AUC improvements of 10.43\% on Taobao, 10.2\% on QB-video, and 5\% on QK-article. Regarding the Logloss metric, \model \ achieves average effective gains of 2.93\%, 6.57\%, and 18.15\% respectively. 
These results further validate the consistency and stability of \model \ superiority across different runs.

\subsection{Performance Comparison with STL Models}\label{app:slt_result}
To further validate the effectiveness of \model, we conducted experiments using three STL models (DeepFM, DIN, and DIEN) across the three datasets. The results, as presented in Table~\ref{tab:STL_result}, demonstrate that our model outperforms these STL baselines in both AUC and Logloss metrics, reinforcing its superiority.

\begin{table}[!t]
  \centering
  \caption{Comparison results with the STL models.}
  \label{tab:STL_result}
  \resizebox{\linewidth}{!}{
    \begin{tabular}{ccc|ccc|cc}
    \toprule
    \multicolumn{3}{c|}{Models} & DeepFM & DIN & DIEN & \model & \textit{Improv.(\%)}  \\

    \midrule
    
    \multicolumn{1}{c}{\multirow{8}[8]{*}{\rotatebox{90}{Taobao}}} 
    & \multicolumn{1}{c}{\multirow{2}[2]{*}{Click}} 
    & AUC      & 0.5501±0.0252 & 0.5632±0.0004 & \underline{0.5644}±0.0003 & \textbf{0.6125}±0.0035 & \textbf{8.52\%} \\ &       
    & Logloss  & 0.6280±0.0695 & \underline{0.3743}±0.0010 & 0.3753±0.0011 & \textbf{0.3632}±0.0015 & \textbf{2.97\%} \\

    \cmidrule{2-8}          
    & \multicolumn{1}{c}{\multirow{2}[2]{*}{Favor}} 
    & AUC      & 0.5810±0.0494 & 0.5721±0.0006 & \underline{0.5861}±0.0004 & \textbf{0.7961}±0.0033 & \textbf{35.83\%} \\ &       
    & Logloss  & 0.2235±0.0218 & 0.1527±0.0006 & \underline{0.1523}±0.0001 & \textbf{0.1321}±0.0015 & \textbf{13.26\%} \\
    
    \cmidrule{2-8}          
    & \multicolumn{1}{c}{\multirow{2}[2]{*}{Cart}} 
    & AUC      & 0.5432±0.0218 & 0.5660±0.0009 & \underline{0.5670}±0.0010 & \textbf{0.6647}±0.0014 & \textbf{17.23\%}  \\ &       
    & Logloss  & 0.3705±0.0511 & 0.2463±0.0007 & \underline{0.2462}±0.0003 & \textbf{0.2359}±0.0004 & \textbf{4.18\%}  \\
    
    \cmidrule{2-8}         
    & \multicolumn{1}{c}{\multirow{2}[2]{*}{Buy}} 
    & AUC      & 0.5655±0.0291 & \underline{0.6191}±0.0014 & 0.6187±0.0007 & \textbf{0.7012}±0.0077 & \textbf{13.26\%}  \\ &       
    & Logloss  & 0.1401±0.0163 & \underline{0.0942}±0.0008 & 0.0946±0.0006 & \textbf{0.0915}±0.0005 & \textbf{2.87\%}  \\

    \cmidrule{2-8}         
    & \multicolumn{1}{c}{\multirow{2}[2]{*}{Average}} 
    & AUC      & 0.5600±0.0314 & 0.5801±0.0005 & \underline{0.5840}±0.0006 & \textbf{0.6936}±0.0040 & \textbf{18.77\%} \\ &       
    & Logloss  & 0.3405±0.0397 & \underline{0.2168}±0.0006 & 0.2171±0.0005 & \textbf{0.2057}±0.0010 & \textbf{5.12\%}  \\

    \midrule

    \multicolumn{1}{c}{\multirow{8}[8]{*}{\rotatebox{90}{QB-video}}} 
    & \multicolumn{1}{c}{\multirow{2}[2]{*}{Click}} 
    & AUC      & 0.6948±0.1238 & 0.8010±0.0029 & \underline{0.8013}±0.0014 & \textbf{0.8497}±0.0070 & \textbf{6.04\%} \\ &       
    & Logloss  & 0.7279±0.4617 & \underline{0.4419}±0.0030 & 0.4427±0.0021 & \textbf{0.4037}±0.0067 & \textbf{8.64\%} \\
    
    \cmidrule{2-8}          
    & \multicolumn{1}{c}{\multirow{2}[2]{*}{Like}} 
    & AUC      & 0.6177±0.1341 & 0.6631±0.0026 & \underline{0.6640}±0.0021 & \textbf{0.8431}±0.0090 & \textbf{26.97\%} \\ &       
    & Logloss  & 0.1633±0.1037 & \underline{0.0406}±0.0001 & \underline{0.0406}±0.0001 & \textbf{0.0310}±0.0012 & \textbf{23.65\%} \\
    
    \cmidrule{2-8}          
    & \multicolumn{1}{c}{\multirow{2}[2]{*}{Share}} 
    & AUC      & \underline{0.5651}±0.0443 & 0.5571±0.0032 & 0.5612±0.0027 & \textbf{0.6630}±0.0189 & \textbf{17.32\%} \\ &       
    & Logloss  & 0.0267±0.0144 & 0.0084±0.0006 & \underline{0.0081}±0.0008 & \textbf{0.0080}±0.0001 & \textbf{1.23\%} \\
    
    \cmidrule{2-8}         
    & \multicolumn{1}{c}{\multirow{2}[2]{*}{Follow}} 
    & AUC      & 0.6169±0.0691 & 0.6130±0.0033 & \underline{0.6271}±0.0022 & \textbf{0.7301}±0.0107 & \textbf{16.42\%} \\ &       
    & Logloss  & 0.0234±0.0125 & 0.0079±0.0004 & \underline{0.0074}±0.0006 & \textbf{0.0060}±0.0001 & \textbf{18.92\%} \\

    \cmidrule{2-8}         
    & \multicolumn{1}{c}{\multirow{2}[2]{*}{Average}} 
    & AUC      & 0.6236±0.0928 & 0.6585±0.0030 & \underline{0.6634}±0.0021 & \textbf{0.7715}±0.0114 & \textbf{16.29\%} \\ &       
    & Logloss  & 0.2353±0.1481 & \underline{0.1247}±0.0011 & \underline{0.1247}±0.0009 & \textbf{0.1124}±0.0020 & \textbf{9.86\%}  \\

    \midrule

    \multicolumn{1}{c}{\multirow{8}[8]{*}{\rotatebox{90}{QK-article}}} 
    & \multicolumn{1}{c}{\multirow{2}[2]{*}{Read}} 
    & AUC      & 0.6013±0.1241 & \underline{0.7817}±0.0044 & 0.7810±0.0045 & \textbf{0.7937}±0.0078 & \textbf{1.54\%} \\ &       
    & Logloss  & 0.7391±0.5120 & \underline{0.1012}±0.0061 & 0.1022±0.0059 & \textbf{0.0921}±0.0060 & \textbf{8.99\%} \\
    
    \cmidrule{2-8}          
    & \multicolumn{1}{c}{\multirow{2}[2]{*}{Like}} 
    & AUC      & 0.6157±0.1505 & 0.8583±0.0080 & \underline{0.8586}±0.0075 & \textbf{0.9250}±0.0074 & \textbf{7.73\%} \\ &       
    & Logloss  & 0.4124±0.2750 & 0.0668±0.0023 & \underline{0.0662}±0.0021 & \textbf{0.0532}±0.0023 & \textbf{19.64\%} \\
    
    \cmidrule{2-8}          
    & \multicolumn{1}{c}{\multirow{2}[2]{*}{Favor}} 
    & AUC      & 0.6198±0.1447 & 0.8571±0.0087 & \underline{0.8610}±0.0082 & \textbf{0.9351}±0.0070 & \textbf{8.61\%} \\ &       
    & Logloss  & 0.1655±0.1061 & 0.0331±0.0021 & \underline{0.0327}±0.0020 & \textbf{0.0251}±0.0004 & \textbf{23.24\%} \\
    
    \cmidrule{2-8}         
    & \multicolumn{1}{c}{\multirow{2}[2]{*}{Follow}} 
    & AUC      & 0.6046±0.1062 & 0.7765±0.0105 & \underline{0.7777}±0.0100 & \textbf{0.8225}±0.0085 & \textbf{5.76\%} \\ &       
    & Logloss  & 0.0252±0.0138 & \underline{0.0073}±0.0087 & 0.0074±0.0085 & \textbf{0.0066}±0.0001 & \textbf{9.59\%} \\

    \cmidrule{2-8}         
    & \multicolumn{1}{c}{\multirow{2}[2]{*}{Average}} 
    & AUC      & 0.6104±0.1314 & 0.8184±0.0079 & \underline{0.8195}±0.0076 & \textbf{0.8691}±0.0070 & \textbf{6.05\%} \\ &       
    & Logloss  & 0.3356±0.2267 & \underline{0.0521}±0.0048 & \underline{0.0521}±0.0046 & \textbf{0.0442}±0.0022 & \textbf{15.16\%} \\

    \bottomrule
    \end{tabular}%
    }
  \label{tab:STL_result}%
\end{table}

\subsection{Complexity Analysis}\label{app:complexity}
We further compare the complexity with the graph-based and the expert-based methods to further validate the efficiency of \model.

\textbf{For the graph-based methods,} for example, MoGENet, given the number of nodes $m+n$, the number of layers $L$ and the feature dimension $d$, the approximate complexity for feature passing in one layer is $O((m+n)d^2)$ and the node feature transformation is $O((m+n)d)$. The overall complexity for $L$ layers is $O(L(m+n)(1+d)d)$. When $m+n$ become extremely large, this complexity significantly raised.
\textbf{For the expert-based methods} (e.g., PLE and STEM), let $h_{e}$, $h_{s}$, $h_{g}$, and $h_{o}$ be the number of shared expert neurons, task-specific expert neurons, gate networks neurons, and tower neurons respectively. The complexity of expert layer is approximately $O(d(h_e + h_g + h_s))$. For a $L$ expert layers method, the complexity is $O(Ld(h_e + h_g + h_s) +  h_e h_o)$ as the number of tasks/shared expert/task-specific expert is a constant.
\textbf{For \model,} let $\widetilde{m}+\widetilde{n}$ be the number of macro nodes, the complexity of each macro graph expert is $O((\widetilde{m}+\widetilde{n})^2d)$. Since there are $T + 1$ modules, the total complexity is $O((T + 1)(\widetilde{m}+\widetilde{n})^2d)$. The macro and micro fusion are composed of CGC layers so the complexity is also $O(Ld(h_e + h_g + h_s))$. The overall complexities of \model \ is approximately $O((\widetilde{m}+\widetilde{n})^2d + Ld(h_e + h_g + h_s) + h_{e}h_o)$. Since the number of macro nodes $\widetilde{m}+\widetilde{n}$ is relatively small, the complexity of \model \ is comparable to that of the expert-based methods.

Overall, \model \ strikes a better balance between computational efficiency and representational power compared to them.

\subsection{Pseudocode}
The complete procedures of \model \ are outlined in Algorithm~\ref{alg:mgoe}.

\begin{algorithm}[htbp]
\caption{\model}
\label{alg:mgoe}
\SetKwInOut{Input}{Input}
\SetKwInOut{Output}{Output}

\Input{Task set $\mathcal{T}$; Task-specific dataset $\mathcal{D}_t$}
\Output{Predicted labels}
\textbf{Macro Task Merging Graph}

Merge interaction matrices into $\mS$\ using Equation~\ref{eq:scoring_function}\;
Construct macro nodes and macro edges by Equation~\ref{eq:macro_process} and Equation~\ref{eq:macro_edge_weight}\;
Build macro task table $T$\;

\textbf{Macro Graph of Experts}

Initialize model parameters $\Theta$\;
\For{each mini-batch}{
    Generate micro user/item embeddings $\ve_u$, $\ve_i$ and macro embeddings $\widetilde{\mE}(\mathcal{\widetilde{N}}_u)$, $\widetilde{\mE}(\mathcal{\widetilde{N}}_i)$\;
    \ForEach{$ t \in \mathcal{T} $}{
        Compute macro weights $w$ and aggregated macro embeddings $\widetilde{\mZ}$ via Equation~\ref{eq:macro_weight} and Equation~\ref{eq:agg_macro_embedding}\;
        Calculate macro graph expert output $\operatorname{MGE}_t$  by Equation~\ref{eq:MGE}\;
        Compute micro neighbor representations $e_\text{nbr}^t$ using Equation~\ref{eq:micro_nbr_agg}\;
        Concatenate micro and macro representations and pass through expert layer $h_t$ by Equation~\ref{eq:E_cat} and Equation~\ref{eq:expert_layer}\;
    }
    Compute final prediction $\hat{\mathrm{Y}}_{ui}$ by Equation~\ref{eq:casading_readout} and Equation~\ref{eq:final_output}\;
    Update model parameters $\Theta$ via $\mathcal{L}_\text{MGOE}$\;
}
\Return Predicted labels\;
\end{algorithm}

\end{document}